\documentclass[graybox]{svmult}

\usepackage{mathptmx}       
\usepackage{helvet}         
\usepackage{courier}        
\usepackage{type1cm}        
\usepackage{makeidx}         
\usepackage{graphicx}        
\usepackage{multicol}        
\usepackage[bottom]{footmisc}
\usepackage{hyperref}        
\usepackage{soul}            
\hypersetup{colorlinks=true,urlcolor=blue}
\usepackage[square,numbers]{natbib}
                       
\usepackage{amsmath}

\begin{document}
\title*{Repeated Bursts: Gravitational Waves from Highly Eccentric Binaries}
\author{Nicholas Loutrel \thanks{corresponding author}}
\institute{Nicholas Loutrel \at Department of Physics, Princeton University, Princeton, NJ 08544 \email{nloutrel@princeton.edu}}
%
%
\maketitle
%
\abstract{
Compact object binaries formed from dynamics interactions will generically have non-zero orbital eccentricity. The gravitational waves from such binaries can change drastically depending on how large the eccentricity is, ranging from emitted in small a subset of orbital harmonics at low eccentricities, to being concentrated into intense bursts of radiation from each pericenter passage at large eccentricities. Gravitational waves from such highly eccentric binaries present themselves an intriguing systems for probing fundamental physics, but also present interesting challenges in terms of detection. The presence of orbital eccentricity in the gravitational wave signature gives an unequivocal method of determining the origin of the binary, while the highly dynamical nature of pericenter passage often enhances the physics associated with matter and gravity. The generation of faithful models of the repeated burst signals has proven challenging, and often sub-optimal detection strategies like power stacking are considered for detection. This chapter reviews our current understanding of eccentric binaries, the leading efforts to model their gravitational wave emission, and the physics that can be probed with these detections.}
\section{Keywords} 

Gravitational Wave Bursts, Eccentric Binaries, Post-Newtonian Approximation, Tests of General Relativity, Binary Neutron Stars, Dynamical Tides, f-modes

\section{Introduction}

The advent of the era of gravitational wave astronomy has seen a remarkable number of important developments in our understanding of compact objects. From the first detection of binary black holes~\cite{Abbott:2016blz}, along with constraints on the dynamical strong field regime of gravity~\cite{Yunes:2016jcc}, to the detection of binary neutron stars~\cite{TheLIGOScientific:2017qsa} that allowed for constraints on the nuclear matter equation of state and astronomical observations revealing the origins of r-process heavy elements within the universe~\cite{Tanvir:2017pws}, the ground-based detectors of the LIGO~\cite{aLIGO} and Virgo~\cite{Virgo} observatories have opened a new window through which to elucidate our universe.

Despite the detection of multiple compact object binaries~\cite{LIGOScientific:2018mvr}, an open question still remains: what is the origin of such systems detected by ground-based observatories? Arguably, the canonical picture of binary formation is that of binary stellar evolution, described in detail in~\cite{Belczynski:2016obo}. Mass transfer and common envelope phases of main sequence stars in binaries can harden the system such that the binary black holes and neutron stars formed as the end state of the binary's supernovae can merge within the lifetime of the universe. A plausible example of such a binary is the {Hulse-Taylor binary}~\cite{Hulse}, comprised of a neutron star and a pulsar, observations of which were used to infer the existence of gravitational waves long before the first detection~\cite{Weisberg}.

Despite it's importance to probing the strong field regime of gravity, there is one particular property of the binary that is worth examining. The binary is actually on a fairly elliptical orbit, with orbital eccentricity of $e\sim0.7$. The binary is approximately three hundred million years from merger, and thus its gravitational wave emission is outside of the detection band of ground based observatories. Nevertheless, one can perform a basic calculation to determine the orbital properties of the binary once it enters their detection band using the quadrupole approximation of gravitational radiation. The backreaction of quadrupole radiation on the orbital dynamics was first calculated by Peters \& Mathews~\cite{Peters:1963ux}, who realized that gravitational radiation generally causes the orbital eccentricity to decrease. A back of the envelope application to the Hulse-Taylor binary reveals that the binary will have negligibly small eccentricity by the time it enters the detection band of ground-based observatories. Indeed, this is the general result of stellar evolution binaries, and all of the current detections are consistent with this prediction.

In recent years, studies of dynamical stellar environments, such as {globular clusters}, have predicted that there is a non-negligible population of compact object binaries that will enter the detection band of ground-based detectors with measurable eccentricity ($e > 0.05$)~\cite{Samsing:2017xmd}. One particular class of binaries of interest in such environments is {gravitational wave capture} binaries, wherein two compact objects on an unbound orbit can form a bound binary through the emission of gravitational waves during the closest approach. Such binaries can be formed with sufficient compactness that their gravitational wave emission is within the frequency band of ground-based observatories, but with a wide range of possible orbital eccentricities, spanning $0.1 < e < 1$~\cite{Rodriguez:2017pec}. A small subset of these binaries will form with large, close to unity ($e\sim 1$) orbital eccentricity. For binary black holes, the merger rate is estimated to be $1-2$ per year per cubic gigaparsec~\cite{Rodriguez:2018pss}, while for binary neutron stars or black hole-neutron star binaries, the rate is significantly lower at 0.04 per year per cubic gigaparsec~\cite{Ye:2019xvf}.

While gravitational wave captures play an important role in determining the population of eccentric binaries of relevance to ground based detectors, there are a host of dynamical interactions at play in globular clusters. Binary-single interactions, wherein a binary system is disrupted by a third compact object, are chaotic in nature, but can settle into quasi-stable configurations comprising an eccentric binary, whose components need not be those of the original binary, orbiting around a third body~\cite{Samsing:2017xmd}. There are two possible outcomes for this scenario. The first is the eccentric binary receives a sufficient kick to its center of mass to eject it from the cluster. In the second, the eccentric binary merges before the third object disrupts it and sends the total system back to a chaotic phase. Both of these scenarios form binaries that will merge within the age of the universe, and are formed in the detection band of space-based observatories with non-negligible eccentricity. If the masses of the binary components are small enough, these binaries will merge within the detection band of ground-based observatories, but with eccentricities too small to be detected, making it difficult to distinguish them from stellar evolution binaries. This highlights the central challenge in determining the origin of the systems observed by LIGO and Virgo.

Another possible channel to form eccentric binaries is galactic nuclei, wherein many of the same dynamical interactions take place as in globular clusters. Binary-single and single-single interactions within active galactic nuclei (AGN) lead to distributions tilted towards higher masses and higher eccentricities~\cite{Samsing:2020tda, Tagawa:2020jnc}. In the case of gravitational wave capture events, $\sim 26-50\%$ of binaries will form at peak gravitational wave frequencies $>10$ Hz with eccentricities $>0.95$, making them the most eccentric source population for ground-based detectors to date~\cite{Gondan:2020svr}. In addition to these dynamical interactions, the orbit of a binary system may be perturbed by the gravitational interaction of the supermassive black hole at the center of the galaxy. Such systems are called {hierarchical triple systems}~\cite{Antognini:2013lpa}, due to there being a separation of scales, specifically the orbital period of the binary is much shorter than the period associated with motion around the third body. In such systems, the effect of the third object causes a trade-off between the mutual inclination angle of the orbits with the orbital eccentricity of the inner binary. Under exceptional circumstances, a resonant state can be achieved via the {Kozai-Lidov}~\cite{Seto:2013wwa} or post-Newtonian effects~\cite{Naoz:2012bx}, wherein the eccentricity can be enhanced by orders of magnitude and become close to unity, i.e. the unbound limit. From simulations of dynamical stellar environments, it is known that triple systems are a common possibility, but it is expected that resonant triples are not likely to occur in nature.

This chapter reviews our current understanding of the gravitational waves from eccentric systems, with particular emphasis on the high eccentricity limit. Within will be explored the fundamentals of the orbital dynamics and gravitational wave emission of eccentric binaries, the methods of detecting gravitational waves from eccentric systems, the approximations and methods used to model the waveforms from binary black holes, and the physics that can be probed with eccentric systems with emphasis on tests of general relativity and neutron star physics. Units where $G = c =1$ are used throughout this chapter.

\section{Eccentric Dynamics}
\label{dyn}

Before we proceed, it is useful to have a clear picture of the dynamics of eccentric binaries and the gravitational wave signals that they generate. A basic understanding of these can be obtained by studying eccentric binaries at so called leading {post-Newtonian}~\cite{PW} order, where the conservative orbital dynamics of the binary are governed by Newton's law of gravitation and the dissipative radiation reaction is controlled by quadrupole radiation. 

We begin with the orbital dynamics, where the equations of motion are
\begin{equation}
\vec{a} = -\frac{M}{r^{2}} \vec{n}
\end{equation}
where $M = m_{1} + m_{2}$ is the total mass of the binary, $r$ is the relative separation of the two bodies, $\vec{a}$ is the relative acceleration, and $\vec{n}$ is the unit normal to the radial separation. The motion of the binary is restricted to a plane, which we choose to be the $x-y$ plane, with the $z$ coordinate orthogonal to the orbit. With this choice $\vec{r} = r \vec{n}$, with $\vec{n} = [\cos\phi, \sin\phi, 0]$ and $(r,\phi)$ are functions of time. Then, $\vec{a} = d^{2}\vec{r}/dt^{2}$, and the equations of motion separate into
\begin{align}
\label{eq:r-eq}
\frac{d^{2}r}{dt^{2}} - r \left(\frac{d\phi}{dt}\right)^{2} &= - \frac{M}{r^{2}}\,,
\\
\frac{d}{dt}\left(r^{2} \frac{d\phi}{dt}\right) &= 0\,.
\end{align}
There are two constants of motion associated with these equations. The first arises from the latter of these equations, which implies that $r^{2}d\phi/dt = h = \text{constant}$, which is recognized as the orbital angular momentum. The second, namely the orbital energy, is found by realizing that the above equations are conserved under a constant shift of the time coordinate, specifically
\begin{equation}
\label{eq:epsilon-eq}
\epsilon = \frac{1}{2} \left(\frac{dr}{dt}\right)^{2} + \frac{h^{2}}{2 r^{2}} - \frac{M}{r^{2}}\,.
\end{equation}
The existence of these conserved quantities allows us to directly integrate the equations of motions. Making use of the orbital angular momentum, Eq.~\eqref{eq:r-eq} can be written purely in terms of $r$ and its time derivatives, specifically
\begin{equation}
\frac{d^{2}r}{dt^{2}} + \frac{h^{2}}{r^{3}} = -\frac{M}{r^{2}}\,.
\end{equation}
Direct integration of this equation once with respect to time gives Eq.~\eqref{eq:epsilon-eq}. The solution to this equation is
\begin{equation}
r = \frac{p}{1 + e \cos\phi}
\end{equation}
where $p$ is the semi-latus rectum of the orbit, and $e$ is the orbital eccentricity. This is supplemented by an evolution equation for the orbital phase $\phi$ derived from the definition of the orbital angular momentum, specifically
\begin{equation}
\label{eq:phidot}
\frac{d\phi}{dt} = \left(\frac{M}{p^{3}}\right)^{1/2} (1 + e \cos \phi)^{2}\,.
\end{equation}
Making use of these solutions, the quantities $(p,e)$ can be related to the orbital energy and angular momentum
\begin{align}
\epsilon &= - \frac{M}{2p}(1 - e^{2})\,,
\\
h &= \left(M p\right)^{1/2}\,.
\end{align}
This completes the solution to the equations of motion.

The geometry of the orbit is controlled by the value of the {eccentricity} $e$. For $e=0$, the radial coordinate $r$ becomes a constant, and the orbit is a circle. For $0<e<1$, the orbit is an ellipse. In an effective one body frame, a small mass $\mu = m_{1} m_{2}/M$ orbits around a large mass $M$ centered at the focus of the ellipse. Closest approach of $\mu$ to $M$ is referred to as pericenter, and the orbital velocity (called the pericenter velocity at this point) is at its greatest value. For $e=1$, the orbit is a parabola, and the pericenter velocity is now the escape velocity. In this case, the binary still sweeps out a total phase $\Delta \phi = 2\pi$, but the orbit extends out to spatial infinity, and as a result, the orbital period is infinite. Finally, for $e>1$, the orbit is a hyperbola and no longer closes. Instead, the orbital sweeps through $\Delta \phi = 2 \arccos(1/e)$.

While the parameterization of the orbit in terms of $\phi$ provides a complete solution to the orbital dynamics, it unfortunately does not admit an explicit closed-form solution in terms of time due to the nonlinear nature of Eq.~\eqref{eq:phidot}. To provide more complete solutions in terms of time, alternative parameterizations are often employed. The starting point is to define the eccentric anomaly $u$, which is related to the orbital phase $\phi$ through
\begin{equation}
\tan\left(\frac{\phi}{2}\right) = \left(\frac{1+e}{1-e}\right)^{1/2} \tan\left(\frac{u}{2}\right)\,.
\end{equation}
Combining this with Eq.~\eqref{eq:phidot}, one obtains the evolution equation for $u$, which can be directly integrated to give {Kepler's equation},
\begin{equation}
\label{eq:kepler}
\ell = u - e \sin u\,,
\end{equation}
where $\ell = (2\pi/T_{\rm orb}) (t-t_{p})$ is the mean anomaly, with $T_{\rm orb}$ the orbital period of the binary and $t_{p}$ the time of pericenter passage. While Kepler's equation provides a useful relationship between the eccentric anomaly and time, it is unfortunately transcendental and has no known closed-form solutions. The main usefulness of Kepler's equation is for the {Fourier decomposition} of the Newtonian two-body problem, which allows one to write relevant quantities in terms of summations on harmonics of the orbital period. For example,
\begin{align}
\label{eq:cos-fourier}
\cos\phi &= - e + \frac{2}{e} (1-e^{2}) \sum_{k=1}^{\infty} J_{k}(k e) \cos(k\ell)\,,
\\
\label{eq:sin-fourier}
\sin\phi &= 2 (1-e^{2})^{1/2} \sum_{k=1}^{\infty} J'_{k}(k e) \sin(k\ell)\,,
\end{align}
where $J_{k}(x)$ is the Bessel function of the first kind and the prime corresponds to differentiation with respect to the argument. For small eccentricities $e\ll1$, $J_{k}(k e) \sim e^{k}$, and the summations can be truncated at a finite number of terms while retaining accuracy compared to a numerical solution to Eq.~\eqref{eq:kepler}. This completes the discussion of the conservative dynamics of the binary.

At leading PN order, the gravitational waves from the binary are controlled by the {quadrupole approximation}, where the metric perturbation takes the form
\begin{equation}
\label{eq:hij}
h_{ij} = \frac{2}{D_{L}} \frac{d^{2}I_{ij}}{dt^{2}}\,,
\end{equation}
where $D_{L}$ is the luminosity distance to the source, and $I_{ij}$ is the quadrupole moment associated with the binary's orbital motion. The observable gravitational wave is given by the transverse traceless (TT) projection of Eq.~\eqref{eq:hij}. Defining $\vec{N}$ to be the line of sight from the observer to the source, specifically
\begin{equation}
\vec{N} = [\sin\iota \cos\beta, \sin\iota \sin\beta, \cos\iota]
\end{equation}
where $\iota$ is the inclination angle of the source and $\beta$ is the angle between the $x$-axis of the orbital frame with the principle axis defining the plus polarization, the TT projection is defined by
\begin{equation}
h_{\rm TT}^{ij} = {P^{ij}}_{mn} h^{mn}
\end{equation}
with the projector
\begin{align}
P^{ij} &= \delta^{ij} - N^{i}N^{j}\,,
\\
{P^{ij}}_{km} &= {P^{i}}_{k} {P^{j}}_{m} - \frac{1}{2} P^{ij} P_{km}
\end{align}
The plus and cross polarizations $(h_{+},h_{\times})$ of the waveform can be defined by constructing the vectors orthogonal to $\vec{N}$. The end result for eccentric binaries is
\allowdisplaybreaks[4]
\begin{align}
h_{+} &= -\frac{2\mu M}{p D_{L}} \Bigg\{\left(1+\cos^{2}\iota\right) \left[ \cos(2\phi - 2\beta) + \frac{5}{4} e \cos(\phi - 2\beta) + \frac{1}{4} e \cos(3\phi - 2\beta) 
\right.
\nonumber \\
&\left.
+ \frac{1}{2} e^{2} \cos(2\beta)\right] + \frac{1}{2} \sin^{2}\iota \left(e \cos\phi + e^{2}\right)\Bigg\}\,,
\\
h_{\times} &= - \frac{2\mu M}{p D_{L}} \cos\iota \left[2\sin(2\phi-2\beta) + \frac{5}{2} e \sin(\phi - 2\beta) + \frac{1}{2} e \sin(3\phi-2\beta) - e^{2} \sin(2\beta)\right]\,.
\end{align}
The polarizations can be further specified as explicit functions of time using the Fourier series in Eqs.~\eqref{eq:cos-fourier}-\eqref{eq:sin-fourier}, specifically
\begin{equation}
h_{+,\times} = -\frac{2\mu M}{p D_{L}} \sum_{k=1}^{\infty} \left[C_{+,\times}(e,\iota,\beta) \cos(k\ell) + S_{+,\times}(e,\iota,\beta) \sin(k\ell)\right]\,,
\end{equation}
where $C_{+,\times}$ and $S_{+,\times}$ are given in Eqs.~(A15)-(A16) in~\cite{Moreno}.

The gravitational waves carry energy and angular momentum away from the binary, with the energy and angular momentum fluxes governed by~\cite{PW}
\begin{align}
{\cal{P}} &= \frac{1}{32\pi} \oint D_{L}^{2} \frac{dh^{\rm TT}_{ij}}{dt} \frac{dh_{\rm TT}^{ij}}{dt} d\Omega\,,
\\
{\cal{J}}^{i} &= \frac{1}{32\pi} \epsilon^{ijk} \oint D_{L}^{2} \left[h^{\rm TT}_{jm}\frac{d{h^{\rm TT}_{k}}^{\;m}}{dt} - \frac{1}{2} \frac{dh^{mn}_{\rm TT}}{dt} x_{j} \frac{\partial h_{mn}^{\rm TT}}{\partial x^{k}}\right] d\Omega\,,
\end{align}
respectively. To leading order, radiation reaction is governed by the {balance laws}, which directly relate the loss of orbital energy and angular momentum to the fluxes averaged over the wavelength of the gravitational waves. For binary systems, the wavelength of the waves is related to the orbital period, so the average simply becomes the orbit average of the above equations, with the evolution equations for the orbital energy and angular momentum becoming~\cite{Peters:1963ux}
\begin{align}
\frac{dE}{dt} &= - \frac{32}{5} \eta^{2} \left(\frac{M}{p}\right)^{5} \left(1-e^{2}\right)^{3/2} \left(1 + \frac{73}{24} e^{2} + \frac{37}{96} e^{4}\right)\,,
\\
\frac{dL}{dt} &= - \frac{32}{5} \eta^{2} M \left(\frac{M}{p}\right)^{7/2} \left(1 - e^{2}\right)^{3/2} \left(1 + \frac{7}{8} e^{2}\right)\,.
\end{align}
The effect of radiation reaction is now fully specified to leading PN order.

\begin{figure*}
\includegraphics[scale=0.2]{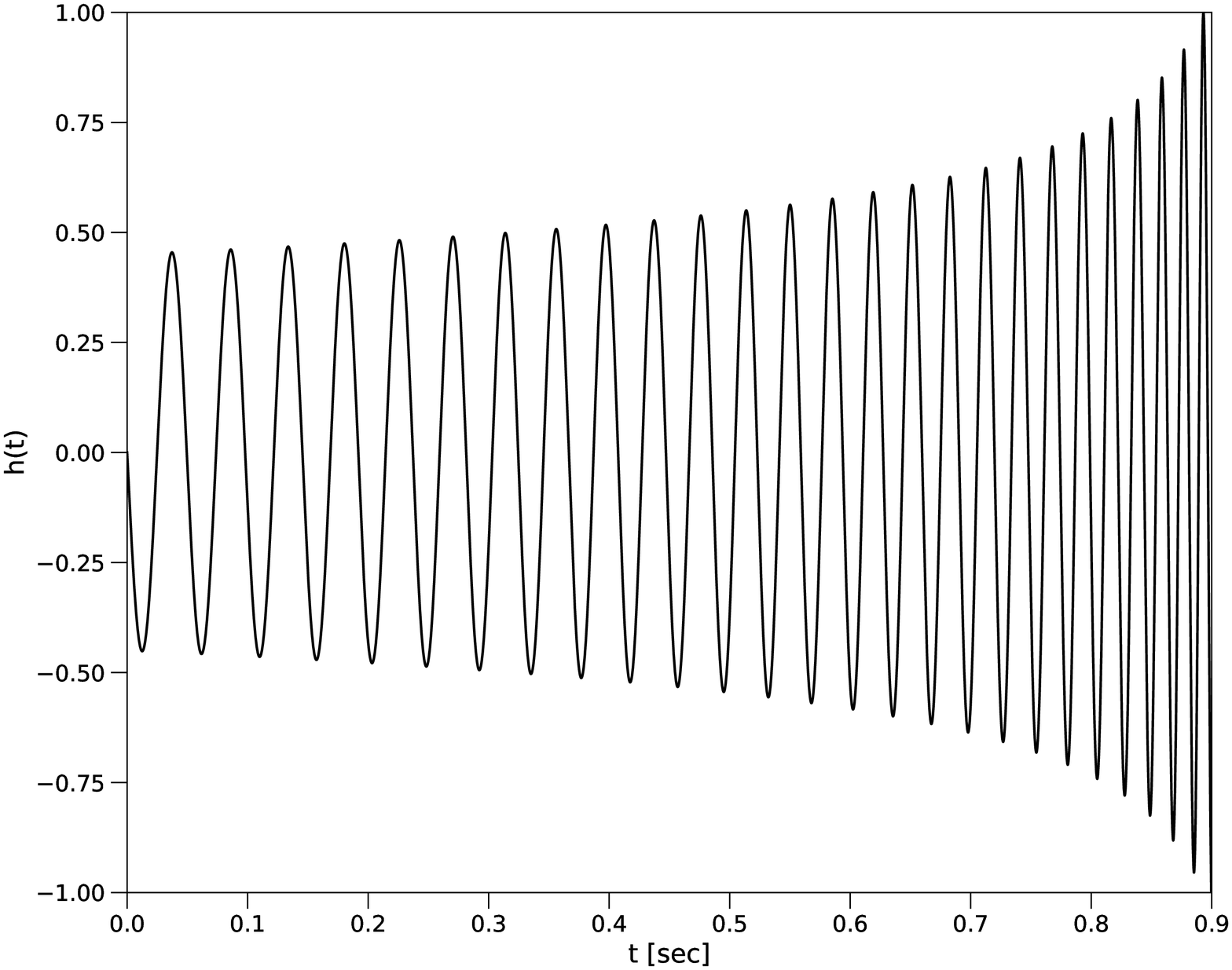}
\includegraphics[scale=0.2]{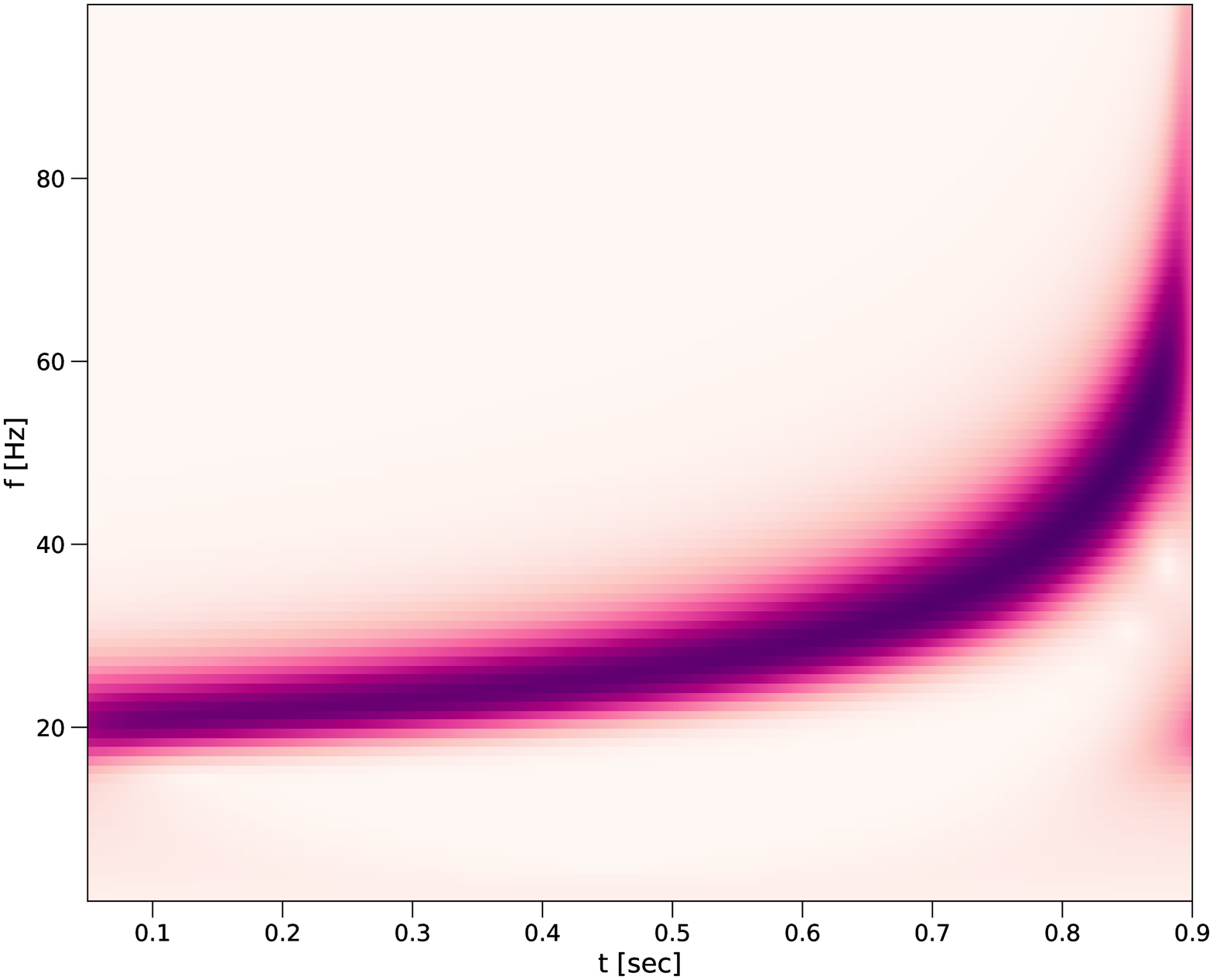}
\includegraphics[scale=0.2]{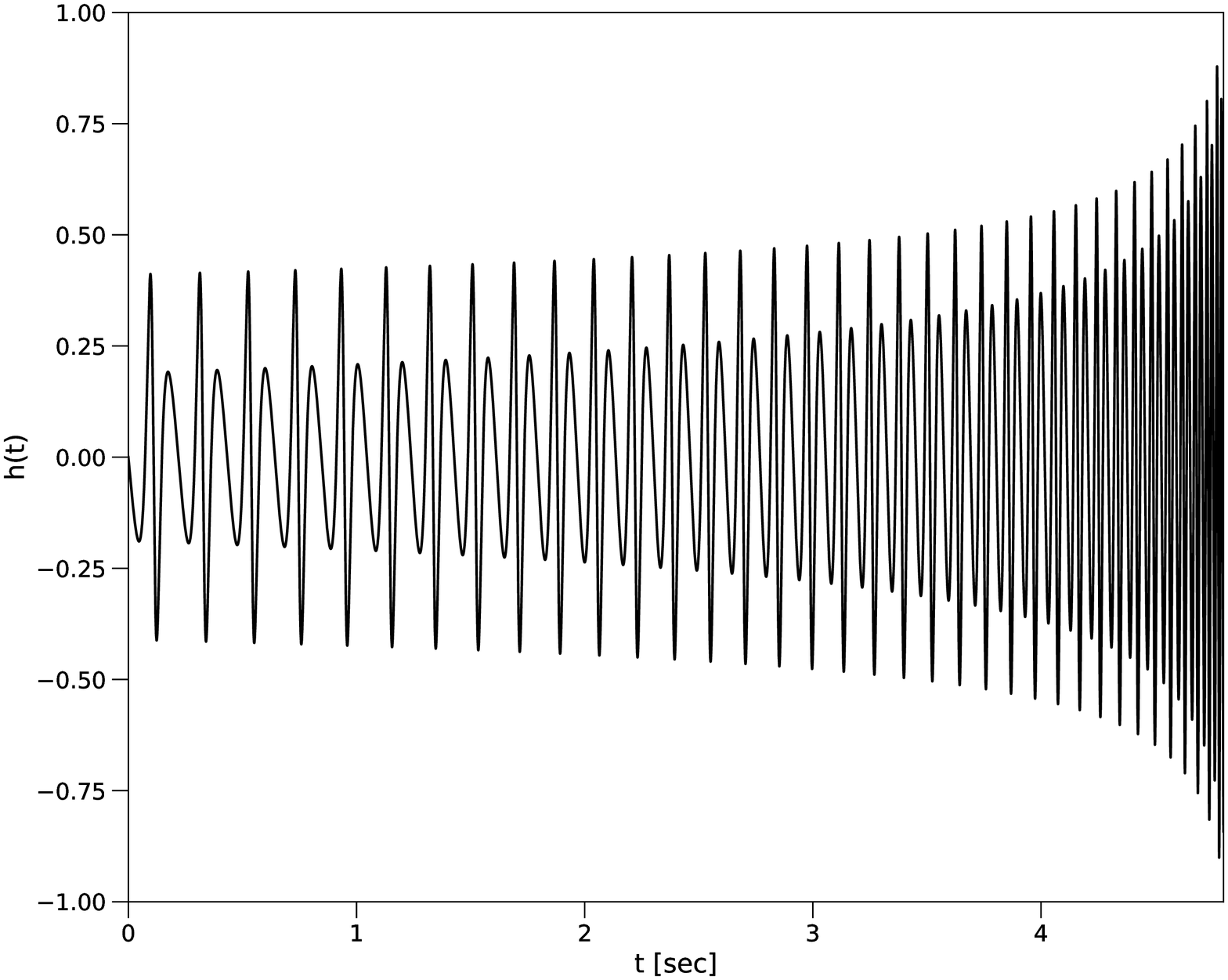}
\includegraphics[scale=0.2]{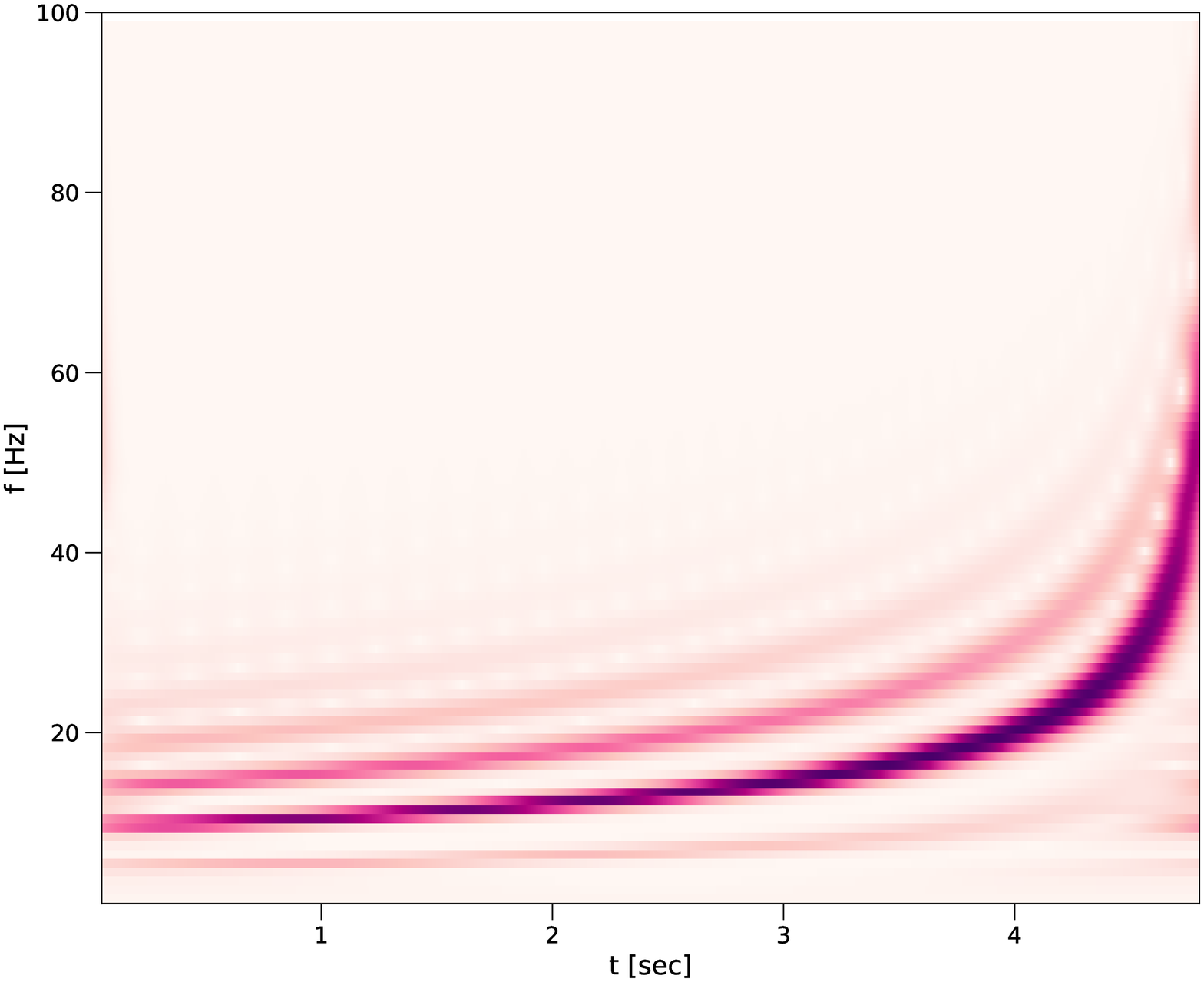}
\includegraphics[scale=0.2]{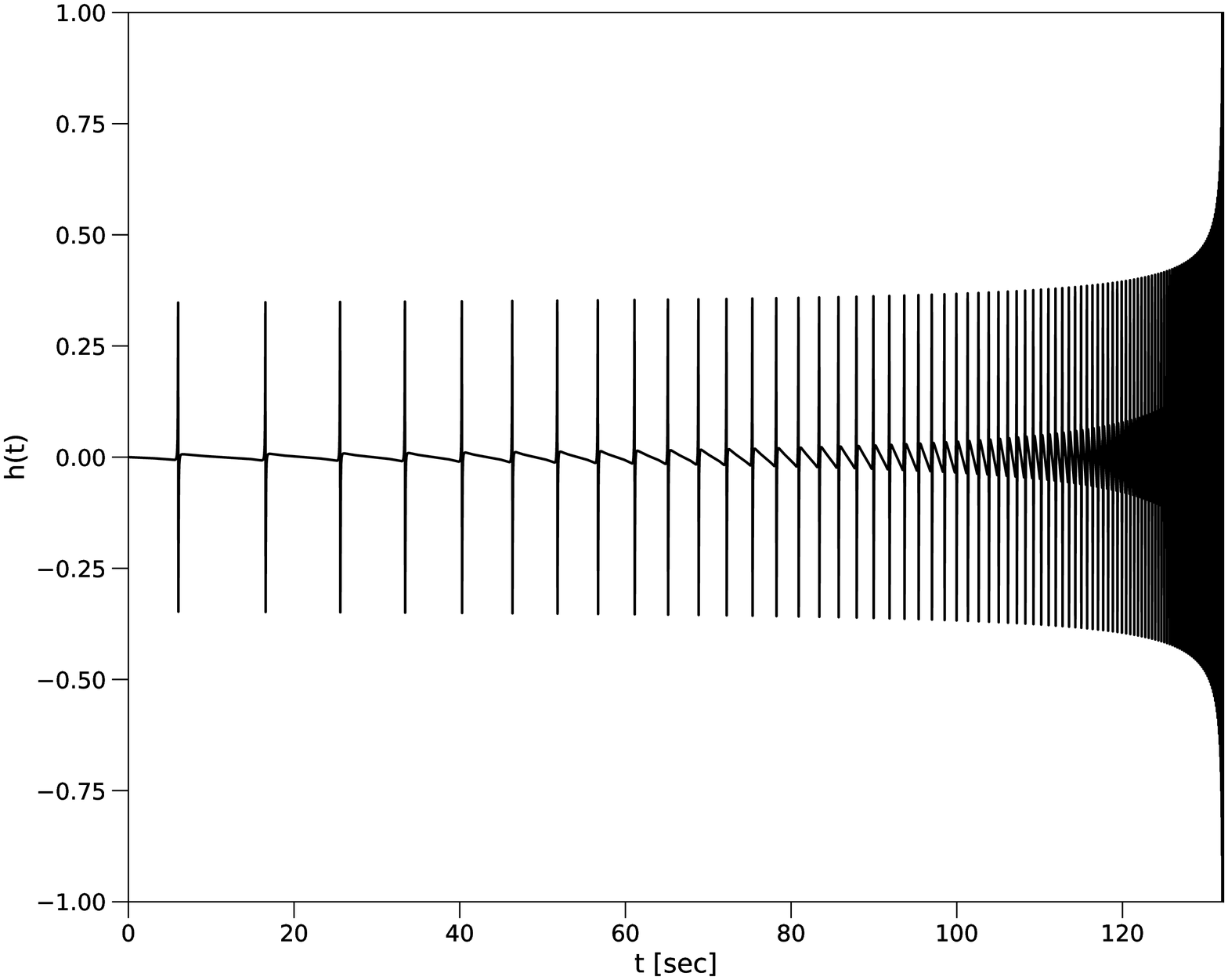}
\includegraphics[scale=0.2]{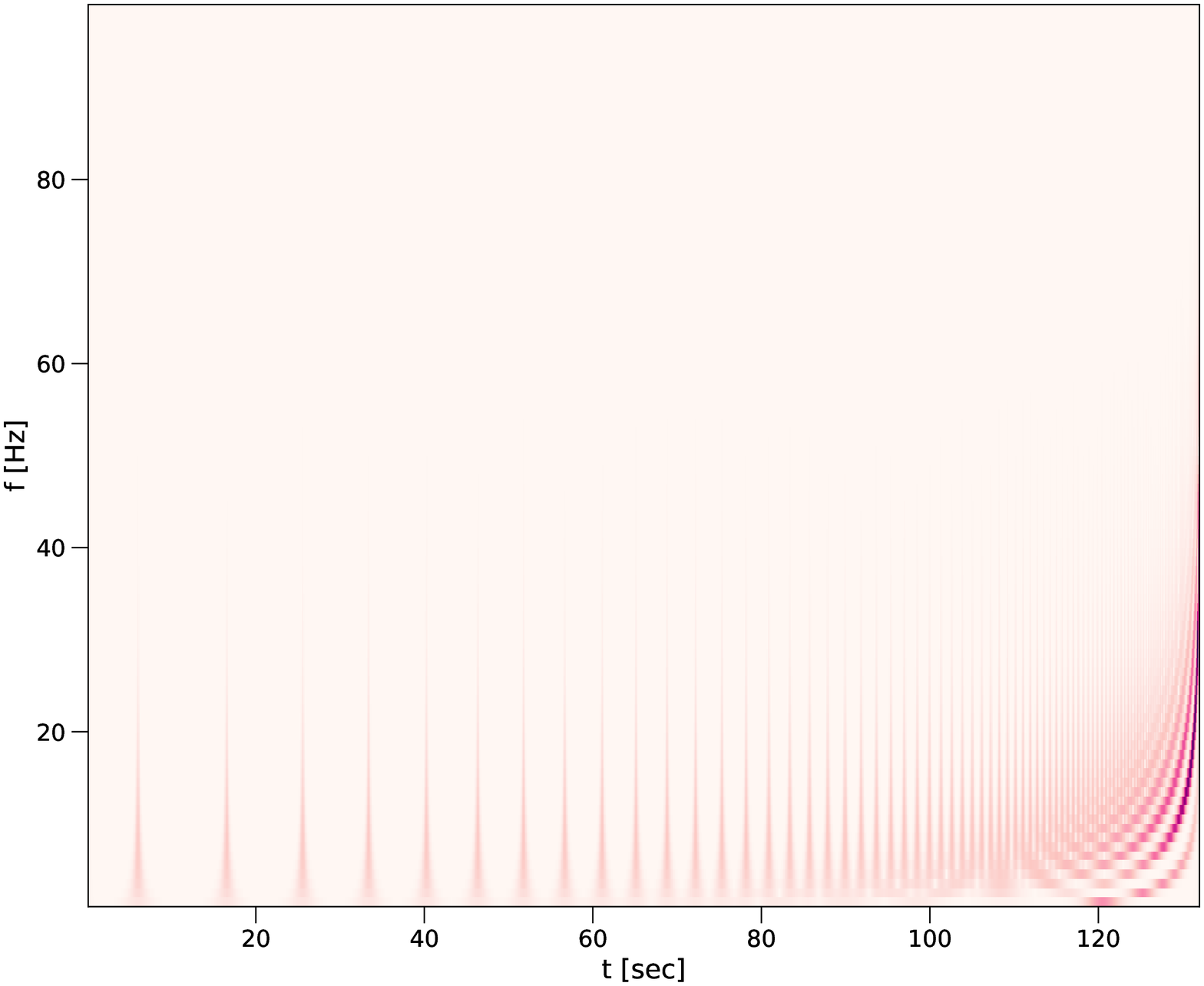}
\caption{\label{hoft} (Left) Plot of the time domain inspiral waveform for a binary with masses $m_{1} = m_{2} = 30 M_{\odot}$, and initial eccentricity $e=0$ (top), $e=0.35$ (middle), and $e=0.95$ (bottom). The waveforms are normalized to their maximum amplitude. As the eccentricity increases, the waveform becomes increasing peaked at pericenter, until it resemble a repeated burst signal. As the binary inspirals, the bursts become louder due to the decreasing of the pericenter distance in each subsequent passage, and occur more rapidly due to the decreasing orbital period. (Right) Q-transform of the inspiral waveforms in the left plots, using wavelets with quality factor $Q=20$. The circular binary (top) only emits power in the second orbital harmonic, while the moderately eccentric binary (middle) emits in multiple harmonics. At high eccentricity (bottom), the power no longer splits into recognizable harmonics and instead becomes localized in time.}
\end{figure*}

Compact object binaries with eccentric orbits have a markedly different gravitational wave signature that their quasi-circular counterparts. {Quasi-circular} ($e\sim0$) binaries emit gravitational waves with frequency twice the orbital frequency, and with a characteristic chirping pattern, becoming louder (higher amplitude) and with increasing frequency as the binary inspirals. This behavior is shown in the time-domain in the top left panel of Fig.~\ref{hoft}. Another common method of analyzing the waveforms from compact binaries is the use of {spectrograms}, a time-frequency representation of the signal. There are a number of ways of achieving these representations, such as a short-time Fourier transform or a wavelet transform. The top right panel of Fig.~\ref{hoft} displays the spectrogram of the corresponding quasi-circular waveform using a Q-transform, a wavelet transform that uses sine-gaussians as the basis function. The spectrogram explicitly shows the chirping behavior of the binary, characterized by the increasing frequency of the signal as a function of time.

For {moderately eccentric} binaries, the behavior of the waveform changes to a deformation of the quasi-circular sequence. As the eccentricity increases, the binary begins emitting at multiple harmonics of the orbital frequency. In the time-domain, the emission of multiple harmonics creates a beating-like behavior in the waveform as can be seen from the middle left panel of Fig.~\ref{hoft}. The explicit emission in multiple harmonics can be seen in the spectrogram in the middle right panel of Fig.~\ref{hoft}. Each harmonic resembles the typical chirping behavior of quasi-circular waveforms.

At sufficiently {large eccentricities} ($e\sim 1$), the waveform changes drastically. The power from gravitational radiation scales as a high power of the orbital velocity (specifically $v^{10}$). For eccentric binaries, the orbital velocity changes throughout the orbit, achieving its maximum at pericenter passage. Thus, as the orbital eccentricity increases, the gravitational radiation becomes more concentrated near pericenter passage. For large orbital eccentricities, the gravitational wave emission no longer resembles the continuous chirping signal of quasi-circular and low eccentricity binaries, but instead becomes a set of discrete bursts emitted during each pericenter passage, as can be seen from the time-domain waveform in the bottom left panel of Fig.~\ref{hoft}. The spectrogram from such a signal, displayed in the bottom right panel of Fig.~\ref{hoft}, reveals that while the bursts are localized within a small time window, they are spread over a broad range of frequencies.

%
\section{Detection strategies}

Having a clear picture in mind of the morphology of eccentric signals, it is useful to discuss the plausible methods through which these signals may be detected before discussing the accuracy of models for binary black holes. This section provides a brief overview of the two common methods described to detect eccentric systems: matched filtering and power stacking.

\subsection{Matched filtering}

{Matched filtering}~\cite{Martel:1999zx} is considered to be the gold standard for detecting any gravitational wave signal. In this method, one develops a phase accurate model for the gravitational wave signal desired to be detected, and then filters the detector data using this waveform model as a kernel. The basic quantity for performing such a search in the noise-weighted inner product, written in the frequency domain as
\begin{equation}
\label{eq:overlap}
(A|B) = 4 \text{Re} \int_{f_{\rm low}}^{f_{\rm high}} df \frac{\tilde{A}(f) \tilde{B}^{\dagger}(f)}{S_{n}(f)},
\end{equation}
where $\tilde{A}(f)$ and $\tilde{B}(f)$ are two frequency-domain signals, $\dagger$ corresponds to complex conjugation, $\text{Re}$ corresponds to the real part, and $S_{n}(f)$ is the noise spectral density of the detector. The limits of integration are chosen to be the lower and upper frequencies that define the detection band of the relevant observatory.

If the waveform model is a perfect representation of the signal being searched for, then the detector's data stream $s$ can be written as $s = h + n$, where $h$ is the waveform model and $n$ is the noise of the detector. Using the above inner product, one can define the match filtered {signal-to-noise ratio} (SNR) as $\rho = (h|h)^{1/2}$. The search is performed by filtering the data $s$ with waveform $h$, and calculating the statistic $\rho$ as a function of time. If the detector noise is stationary and gaussian, one would claim a detection of a signal when $\rho$ achieves a pre-determined threshold. In practice, however, the noise of ground-based detectors in not stationary and gaussian, and typically contains a large number of glitches~\cite{Cabero:2019orq}, spurious noise that creates regions of excess power in spectrograms whose origin may or may not be known. In some cases, glitches may resemble the types of signals being searched for, which is especially true in the case of bursts from highly eccentric systems. When this happens (which is very often for ground-based detectors), it is necessary to calculate the false alarm rate (FAR) of the signal, i.e. how often would a particular signal be created purely by detector noise. The match filtered search would then not just require a threshold SNR, but also a FAR above the threshold determined by the background noise.

How does one determine if a particular waveform model is sufficiently accurate to be used in a match filtered search? This is probed using the {match statistic}, which is also sometimes referred to as the faithfulness or effectualness of the model~\cite{Buonanno:2009zt}. The match is defined as
\begin{equation}
{\cal{M}} = \underset{t_{c}, \phi_{c}}{\text{max}} (h_{\rm exact}|h_{\rm model}(t_{c}, \phi_{c}))
\end{equation}
where $h_{exact}$ is a waveform considered to be an exact representation of what one might expect a signal to look like in nature, and $h_{\rm model}$ is the waveform model being considered, which is dependent on a time $t_{c}$ and phase $\phi_{c}$ shift. The exact procedure for maximizing over these parameters depends on how complicated the model is. In the simplest case, such as a PN quasi-circular waveform, the model can be written as $h_{\rm model} = h_{0}(\mu^{a}) e^{2i\pi f t_{c} + i \phi_{c}}$, where $\mu^{a}$ are the remaining parameters of the model. If the waveform can be written in this manner, then the maximization over $t_{c}$ may be achieved by computing the inverse Fourier transform of the integrand in the inner product, while $\phi_{c}$ can trivially be maximized over after this procedure. 

In general, the match is in the range $0 \le {\cal{M}} \le 1$. Using the analogy between the dot product of two vectors, if ${\cal{M}} = 1$, then the waveforms are said to be perfectly in phase, while if ${\cal{M}} = 0$, then the waveforms are maximally out of phase. The closer the match is to unity, the more accurate a representation the model is to the exact waveform. The match is also related to the loss in number of detections if one used a particular waveform model for a match filtered search. The higher the match, the less signals one would miss is a search. From this notion, a threshold is typically chosen for the waveform model to achieve in order to be used in such a search, specifically ${\cal{M}} \ge 0.97$, which ensures that one misses at most $10\%$ of signals~\cite{Buonanno:2009zt}. The match statistic will be important in the discussion of waveform models for binary black holes in the next Sec.~\ref{bbh}.

\subsection{Power stacking}
\label{stack}

One of the primary challenges of a match filtered search is obtaining models with sufficient phase accuracy compared to the signals that we expect from nature. As we will discuss in the Sec.~\ref{nr}, there are few accurate numerical relativity waveforms for highly eccentric binaries that cover more than one pericenter passage. This means that in the high eccentricity limit, we are currently missing the requisite models to perform a match filtered search. We must then turn our attention to sub-optimal strategies for achieving detections of highly eccentric binaries.

A commonly discussed strategy is that of {power stacking}~\cite{Tai:2014bfa}. In this strategy, one combines the power in multiple signals to boost the SNR in order to achieve the threshold needed for detections. For highly eccentric binaries, these signals are the individual bursts of the inspiral sequence of a single binary. However, power stacking is a general technique that can also be used to combine multiple low significance signals from different systems~\cite{Yang:2017xlf}. Assuming each signal is the same SNR, then the total SNR of the stacked signal scales as $N^{1/4}$, where $N$ is the number of signals. This is sub-optimal in the sense that for match filtering, the enhancement in the SNR scales as $N^{1/2}$ through the stacking of amplitude as opposed to power.

Power stacking for highly eccentric binaries has been considered in the idealized scenario of stationary gaussian detector noise~\cite{Tai:2014bfa}. To simulate the waveforms generated by highly eccentric binaries, the authors made use of the effective Kerr spacetime model explained in detail in Sec.~\ref{kerr}. The analysis carried out therein makes use of wavelet transforms, with basis functions $\psi_{k}$, to characterize the bursts of the simulated signal. From the wavelet transform, one can define two relevant statistics. The first is the sum ${\cal{C}} = \sum_{k=1}^{N} |<\psi_{k}|h'>|^{2}$, where $h'$ is the whitened signal
\begin{equation}
h'(t) = \int_{-\infty}^{\infty} df \frac{\tilde{h}(f)}{\sqrt{S_{n}(f)}} e^{2\pi i f t}\,,
\end{equation}
and $<|>$ defines the time-domain inner product
\begin{equation}
<f|g> = \int_{-\infty}^{\infty} dt f(t) g^{\dagger}(t)
\end{equation}
with $\dagger$ corresponding to complex conjugation. Note that the statistic ${\cal{C}}$ is the sum of power in each wavelet. The N-burst signature is then the set of wavelets ${\cal{S}}_{N} = \{\psi_{1}, ..., \psi_{N}\}$ that maximizes ${\cal{C}}$. With the N-burst signature, one can then compute statistic
\begin{equation}
{\cal{E}}(\tau) = \sum_{i=1}^{N} \underset{\varphi_{i} \in B_{i}}{\text{max}} |<\varphi_{i}(t) | x(t+\tau)>|^{2}
\end{equation}
where $x$ is the detector output and $\varphi$ are the wavelets in the set $B_{i} = \{\varphi | D(\varphi, \psi_{i}) \le \xi_{i}\}$, with $D$ the Euclidean distance between wavelets. In order to achieve detection, one requires ${\cal{E}}$ to achieve a minimum value set by the false alarm rate $N_{\rm false} = \eta(T/\delta t)$, where $T$ is the total observation time and $\delta t$ is determined by the sampling rate. For a false event rate of $1\text{yr}^{-1}$ and step size of $10\text{ms}$, one requires $-\ln\eta \sim 22$, which sets the threshold for ${\cal{E}}$.

Simulated searches were performed using this method through the use of Monte Carlo simulations. The search is generally more efficient at detecting equal mass binaries with higher total masses and smaller pericenter distances, as can be seen from Fig.~17-19 in~\cite{Tai:2014bfa}. The strategy is also significantly more robust to modeling errors in the fluxes of gravitational radiation than matched filtering (see Fig.~20 in~\cite{Tai:2014bfa}). The drawback of this search is that signals generally need to be closer than for matched filtering searches by as much as a factor of 3, which is visualized in Figs.~10-16 in~\cite{Tai:2014bfa}. However, the search is more sensitive than single burst searches.

\section{Modeling of Black Hole Binaries}
\label{bbh}

The detection strategies described in the previous section require one to have models of the types of signals being searched for. For matched filtering, this requires the construction of phase accurate waveforms, while in the case of power stacking, timing models can aid is locating bursts in time-frequency space. This section details some of the relevant models for approximating the gravitational waves from eccentric {binary black holes}.

\subsection{Numerical models}

Generically, the equations governing the conservative and dissipative dynamics of eccentric binaries in general relativity are too complicated to be solved in closed-form. Thus, there are two ways one may proceed to obtain the waveforms from these systems. The first is to simply numerically integrate the equation of motion for the orbital dynamics, which in turn determines the gravitational waveform of the binary. Below is detailed many of the methods and approximations used to achieve this. The second method is to apply analytic approximations to solve the necessary equations, a discussion of which can be found in Sec.~\ref{an}.

\subsubsection{Numerical Relativity}
\label{nr}

Currently, the most accurate waveforms compared to signals in nature are achieved through {numerical relativity}, which seeks to solve the Einstein field equations as an exact system numerically for binary mergers. As a basic concept, the simulations model black holes as punctures on a coordinate grid that defines a slice in time. Initial data constructed from constraint equations is specified on the initial slice, and then evolved according to a set of evolution equations. There are challenges related to eccentric binaries in numerical relativity, and some limited results for eccentric binary black holes that are worth discussing at this stage.

Generic initial data leads to punctures moving on eccentric orbits, which can be seen qualitatively from oscillations in the coordinate radial velocity of the punctures. Historically, the presence of eccentricity has been viewed as a negative aspect of the simulations, with many procedure developed to quantify and reduce the eccentricity~\cite{Pfeiffer:2007yz, Husa:2007rh, Tichy:2010qa}. Primarily, this is a result of the expectation that most binaries of relevance to ground-based detectors will be quasi-circular. However, in recent years the Simulating Extreme Spacetimes (SXS)~\cite{Boyle:2019kee} program has explored binary black holes with small to moderate eccentricity in detail.

The {highly eccentric} limit, on the other hand, poses a challenge for numerical simulations. The main issue is a result of the multiple timescales associated with the problem. The gravitational wave emission is concentrated during pericenter passage, and thus occurs on a small timescale. However, the time between bursts is effectively the orbital period of the binary, which can be orders of magnitude larger than the pericenter passage timescale in the high eccentricity limit. Thus, one must invest sufficient computation resources to integrate over the long orbital timescale, while also retaining sufficient resolution to accurately resolve the dynamics at pericenter, and it is difficult to retain convergence over multiple orbits. The few simulations of the high eccentricity limit typically only resolve a single pericenter passage, except for dynamical captures with sufficiently small impact parameter. These simulations have primarily focused on black hole-neutron star binaries and binary neutron stars, so the results will be presented in Sec.~\ref{ns}.

There are a few results from numerical relativity concerning black holes in highly eccentric orbits that are worth mentioning here. The first is the work of Healy, Levin, \& Shoemaker~\cite{Healy:2009zm}, who investigated the generality of zoom-whirl orbits, wherein the binary ``whirls" around the center of mass in nearly circular motion during closest approach before ``zooming" out along a highly eccentric orbit. Such behavior is well known in extreme mass ratio inspirals (EMRIs), but at the time, it was not known how general this behavior was especially for more comparable masses. It was expected that dissipation due to gravitational radiation would cause the binary to circularize before zoom-whirl behavior could be initiated. Healy, Levin, \& Shoemaker showed that zoom-whirl behavior can be found in highly eccentric binaries of comparable masses along highly eccentric orbits with sufficiently small impact parameters. In general, the zoom-whirl behavior was more prevalent in systems with larger mass ratios and larger spins.

Second, Damour et. al.~\cite{Damour:2014afa} considered the scattering of black holes on hyperbolic-like orbits using numerical relativity, specifically computing the scattering angle $\chi$ as a function of the initial impact parameter. Comparisons therein to predictions of the scattering angle as computed in post-Newtonian theory and the effective one-body framework showed that the post-Newtonian predictions become increasingly inaccurate the smaller the initial impact parameter, while the effective one-body predictions have significantly improved accuracy with errors of at most $\sim 5\%$.

\subsubsection{Effective Kerr spacetime}
\label{kerr}

One of the most important models of highly eccentric inspirals is the effective Kerr spacetime~\cite{East:2012xq}. The model takes some motivation from the effective one-body formalism~\cite{Buonanno:1998gg}, where inspiral dynamics of non-spinning binaries are model as geodesics of a deformed Schwarzschild black hole. Further, zoom-whirl dynamics in equal mass binaries is typically better approximated by geodesics of Kerr spacetimes rather than Schwarzschild spacetimes~\cite{Pretorius:2007jn}. The central idea of the model is that, for small impact parameters, the spacetime resembles a single throat of an effective black hole, with the orbital dynamics governed by geodesics in this spacetime. This approximation is similar to those used in the close-limit approximation~\cite{Price:1994pm,Abrahams:1995wd}.

The {Kerr spacetime} is given by the metric, in Boyer-Lindquist coordinates,
\begin{align}
ds^{2} &= - \left(1 - \frac{2Mr}{\Sigma}\right) dt^{2} + \frac{\Sigma}{\Delta} dr^{2} + \Sigma d\theta^{2} 
\nonumber \\
&+ \left(r^{2} + a^{2} + \frac{2Ma^{2} r \sin^{2}\theta}{\Sigma}\right)\sin^{2}\theta d\phi^{2} - \frac{4Mar \sin^{2}\theta}{\Sigma} dt d\phi\,,
\\
\Sigma &= r^{2} + a^{2} \cos^{2}\theta\,,
\\
\Delta &= r^{2} - 2 M r + a^{2}\,,
\end{align}
where $M$ and $a$ describe the mass and spin of the black hole. In general, $a < M$ for astrophysical sources, with $a=M$ called the extremal limit. The model is described as an effective Kerr black hole since $M$ is taken to be the total mass of the binary and the model sets $a = a_{\rm eff}$, where $a_{\rm eff} = \mu \tilde{L}/M + a_{\rm BH}$, where $\mu$ is the reduced mass of the binary, $\tilde{L}$ is the reduced angular momentum of the geodesics, and $a_{\rm BH}$ is the total spin of the component black holes of the binary. As a result, for large orbital separations $a_{\rm eff} > M$, and the spacetime resembles that of a super-extremal black hole (which is actually a naked singularity). One may worry that this creates possible pathologies within the model. However this only occurs when $\tilde{L}$ is sufficiently large. Assuming a Newtonian mapping $\tilde{L} = (M p)^{1/2}$ where $p$ is the semi-latus rectum of the orbit, $a_{\rm eff} = M$ when $p = (M/\mu^{2}) (M^{2} - 2 a_{\rm BH} M + a_{\rm BH}^{2})$. For equal mass $\mu = (1/4) M$ and non-spinning $a_{\rm BH} = 0$ binaries, this corresponds to $p = 16 M$, which corresponds to an orbit that is outside the region where pathologies exist.

The inspiral phase of the binary coalescence is described by the geodesics of the effective Kerr spacetime, with the geodesics described by the reduced orbital energy $\tilde{E}$ and angular momentum $\tilde{L}$. Radiation reaction is included in a {post-Minkowski}~\cite{PhysRevD.56.826} style approach, where, under the emission of gravitational waves, these quantities are promoted to functions of time through the quadrupole formula
\begin{align}
\frac{d\tilde{E}}{dt} &= - \frac{\mu}{5} \dddot{\cal{I}}_{jk} \dddot{\cal{I}}^{jk}\,,
\\
\frac{d\tilde{L}}{dt} &= - \frac{2\mu}{5} \epsilon^{zij} \ddot{\cal{I}}_{ik} {\dddot{\cal{I}}_{j}}^{k}
\end{align}
where ${\cal{I}}_{jk}$ is the quadrupole moment of the geodesic, and $\epsilon_{ijk}$ is the Levi-Civita symbol. The time domain waveform is then given by
\begin{equation}
h = \frac{2}{D_{L}} \left(\ddot{\cal{I}}_{xx} + i \ddot{\cal{I}}_{xy}\right)\,,
\end{equation}
where $D_{L}$ is the distance to the source. The merger and ringdown of the gravitational wave signal are modeled through the {implicit rotating source} (IRS) model, where the multipolar moments of the gravitational radiation are associated with the mass multipole moments of a rigidly rotating source, specifically the remnant black hole in this case~\cite{Baker:2008mj, Kelly:2011bp}. The phase of the model is provided by a function that asymptotically approaches the dominant quasi-normal mode frequency of the remnant black holes. This phase function, given in Eq.~(6)-(7) in~\cite{East:2012xq}, has three free parameters, plus one for the amplitude, that must be fit to numerical relativity simulations.

In~\cite{East:2012xq}, the model was validate against both post-Newtonian waveforms and numerical relativity fly-by waveforms. For large orbital separations, the loss of energy and angular momentum of the geodesic model converges to that of the 2.5- and 3.5-PN approximations, but these approximations diverge for close pericenter passages, specifically $r_{p} < 10M$. On the other hand, the model was compared to numerical relativity fly-by waveforms using the match statistic. The free parameters of the geodesic model, namely the pericenter distance $r_{p}$ and eccentricity $e$, were fit by finding the highest match. The match was found to be more sensitive to the parameter $r_{p}$ than to the eccentricity of the orbit. The comparison was performed with binary black holes, black hole-neutron star binaries, and binary neutron stars. For the binary black hole case, the match was very high at 0.99, while for the cases with neutron stars, the match was very sensitive to the closest approach. This is due to the fact that the model does not include finite size effects, which become increasingly important as the pericenter distance decreases.

Thus far, the model is designed to describe binary black holes, since it does not include the matter effects necessary to describe binaries with neutron stars. The model has also only been validated with non-spinning binaries, and only for single pericenter passages, due to the lack of fully inspiral waveforms from numerical relativity simulations. Finally, the model has not been validated in a simulated data analysis study.

\subsubsection{Inspiral-merger-ringdown waveforms}

Generically, one cannot solve for the full {inspiral-merger-ringdown} structure of the waveform of binary systems without using numerical relativity. However, even numerical relativity has its limits in terms of the number of gravitational wave cycles it can accurately simulate. On the other hand, the early inspiral evolution where the orbital velocity of the binary is small compared to the speed of light can be accurately modeled by the post-Newtonian approximation. A number of approaches have been used to combine these two approaches to obtain complete waveform that accurately model the full coalescence of binary systems. Two of these, IMRPhenom and SEOBNR models, are standard models employed by the LIGO and Virgo collaborations, while a third, the ENIGMA model, has been developed specifically for eccentric systems.

The {IMRPhenom}~\cite{Ajith:2007qp, Santamaria:2010yb} family of models combines post-Newtonian approximation for the inspiral phase of the coalescence with numerical relativity simulations for the merger-ringdown phase.  In the post-Newtonian approximation, waveforms can be computed analytically in the Fourier domain by application of the stationary phase approximation~\cite{Bender}, and thus the matching is typically done in the frequency domain. The two separate zones, early inspiral and merger-ringdown, are interpolated in between in an intermediate matching zone. The resulting hybrid waveform is then fit to a parameterized model with phenomenological coefficients, that are related to the masses and spins of the binary. While much recent work has gone into extending the IMRPhenom models to include spin precession, there are currently no eccentric IMRPhenom models. However, the first step toward this, namely developing an post-Newtonian eccentric inspiral waveform, has been achieved in~\cite{Moore:2019xkm}, and is detailed in Sec.~\ref{moore}.

The {effective one body} (EOB) approach~\cite{Buonanno:1998gg} takes some inspiration from techniques in quantum field theory. Specifically, the generic two body problem in general relativity is mapped to an effective one-body system describe by a deformed Schwarzschild or Kerr spacetime metric characterized by effective potentials, which are matched to known limits. In particular, much of the free parameters in the potentials are known from the post-Newtonian expansion, while any unknown coefficients are fit by matching to numerical relativity simulations. The potentials also properly reduce to the correct test mass limit of the self-force formalism~\cite{Barack:2018yvs}. The orbital dynamics and gravitational waves are then characterized by geodesics of the effective spacetime. Much of the work on SEOBNR waveforms have focused on quasi-circular and spinning binaries. However, more recently, two versions of these models have been developed for moderately eccentric binaries, and shown to provide good agreement with numerical relativity SXS simulations~\cite{Cao:2017ndf, Liu:2019jpg, Chiaramello:2020ehz}. Finally, Nagar et. al.~\cite{Nagar:2020xsk} have developed an EOB model for dynamical capture binary black holes, which reproduces the scattering angle of numerical relativity hyperbolic encounters to within a few percent. This last model constitutes the only currently available inspiral-merger-ringdown model for highly eccentric binaries.

Finally, the {ENIGMA} model~\cite{Huerta:2016rwp, Huerta:2017kez} makes use of time-domain post-Newtonian approximations of the inspiral dynamics of the binary. However, these approximations are only known up to third post-Newtonian order for generic eccentric systems. The ENIGMA model also encorporates higher post-Newtonian order corrections to the energy flux and binding energy from the self-force formalism up to sixth post-Newtonian order. For the merger-ringdown part of the waveform, the model makes use of Gaussian process emulation, wherein a machine learning algorithm is trained to interpolate between numerical relativity waveforms in a dataset. To match the inspiral and merger-ringdown parts together, the model finds the attachment point by maximizing the overlap, given by Eq.~\eqref{eq:overlap}, between quasi-circular ENIGMA waveforms with SEOBNRv4 waveforms. The model was validated against Einstein Toolkit numerical relativity waveforms up to eccentricity $e=0.2$ ten cycles before merger in~\cite{Huerta:2017kez}.

\subsection{Analytic models}
\label{an}

Complementary to the numerical approaches detailed above, much work has gone into the development of analytic waveform models for eccentric binaries. These models usually only cover the inspiral phase of the coalescence, but have the benefit of typically being faster to evaluate than numerical models. The radiation reaction equations for eccentric binaries are more difficult to solve analytically than their quasi-circular counterparts, and usually one has to employ assumptions and approximations to solve them accurately. Most eccentric waveform models use some version of the {post-circular} expansion~\cite{Yunes:2009yz}, an expansion about low eccentricity. An excellent, short review of these current models can be found in Sec. of~\cite{Moore:2018kvz}. The following sections provide an overview of analytic waveforms models that are tailored to highly eccentric systems.

\subsubsection{Adiabatic waveforms}
\label{moore}

Most inspiral waveforms are usually constructed in the {adiabatic approximation}, where the loss of orbital energy and angular momentum is balanced by the orbit averaged energy and angular momentum fluxes of gravitational radiation. The current state-of-the-art of adiabatic inpsiral waveforms for eccentric binaries is the NeF model of~\cite{Moore:2018kvz}, and it's post-Newtonian extension~\cite{Moore:2019xkm}. 

The main challenge of developing adiabatic eccentric models is to solve for the relationship between the orbital frequency $F$, the mean anomaly $\ell$, and eccentricity $e$. At leading PN order, these quantities evolve under radiation reaction according to
\begin{align}
\label{eq:dedt}
\frac{de}{dt} &= -\frac{304}{15} \frac{\eta}{M}\left(2\pi M F\right)^{8/3} \frac{e\left(1 + \frac{121}{304} e^{2}\right)}{(1-e^{2})^{5/2}}  
\\
\label{eq:dFdt}
\frac{dF}{dt} &= \frac{96}{10\pi} \frac{\eta}{M^{2}} \left(2\pi M F\right)^{11/3} \frac{\left(1 + \frac{73}{24} e^{2} + \frac{37}{96} e^{4}\right)}{(1-e^{2})^{7/2}}
\\
\frac{d\ell}{dt} &= 2\pi F
\end{align}
Generically, these equations can only be solved by change of variables from $t$ to $e$ to obtain $F(e)$, $t(e)$, and $\ell(e)$, which the latter two depend on hypergeometric functions. There are two issues with this. The holy grail of analytic waveforms is to obtain them in the Fourier domain analytically, which can be achieved by application of the stationary phase approximation. In order to do this, one needs then $t(F)$ and the stationary point of the phase gives the relationship between the Fourier frequency $f$ and the orbital frequency. One thus has to invert $F(e)$ to obtain $e(F)$, which is challenging since this is a transcendental equation. The second issue is a result of $t(e)$ and $\ell(e)$ depending on hypergeometric functions, which are typically slow to evaluate numerically.

These problems can be overcome by use of post-circular (low eccentricity) expansions. Comparisons of the low eccentricity inversion of $F(e)$ to numerical inversions reveals that the approximation is well behaved and controllable depending on the order of the expansion. The residuals can be further improved by use of re-summations of the series using Chebyshev and Legendre polynomials. On the other hand, the hypergeometric functions in $t(e)$ and $\ell(e)$ are relatively simple functions of the eccentricity, and are well approximated by sufficiently high order post-circular expansions, which are significantly faster to evaluate. 

With the analytic solutions to the radiation reaction equations in Eqs.~\eqref{eq:dedt}-\eqref{eq:dFdt} in hand, the time-domain waveform can be written analytically through a decomposition in terms of harmonics of the orbital period, specifically
\begin{equation}
h_{+,\times}(t) \sim \sum_{j} A^{(+,\times)}_{j}[e(F), F] e^{ij\ell(F)}
\end{equation}
where $h_{+,\times}$ are the waveform polarizations, $A_{j}$ are amplitude functions written in terms of Bessel functions of the first kind, and $F$ is a function of time through Eq.~\eqref{eq:dFdt}. This decomposition was originally considered in the calculation of adiabatic fluxes by Peters \& Mathews~\cite{Peters:1963ux}, and was later extended to the waveforms by Moreno-Garrido et. al.~\cite{Moreno}. These waveforms can be analytically Fourier transformed through application of the stationary phase condition, with the stationary point given by $j F(t_{\star}) = f$, where $t_{\star}$ is the time associated with the stationary point, $f$ is the Fourier frequency, and $j$ is the harmonic number. The NeF model is then, schematically,
\begin{align}
\tilde{h}_{+,\times}(f) \sim \sum_{j} {\cal{A}}_{j}^{(+,\times)}[e(f), f] e^{i\psi_{j}(f)} 
\end{align}
where ${\cal{A}}_{j}$ are known amplitude functions, and
\begin{align}
\psi_{j}(f) &= j\ell_{c} - 2\pi f t_{c} - \frac{15j}{304\eta} \frac{\sigma(e_{0})^{-5/2}}{\left(2\pi M F_{0}\right)^{5/3}} e(f)^{30/19} I[e(f)]
\end{align}
with $(\ell_{c}, t_{c})$ the phase and time of coalescence, $(e_{0}, F_{0})$ the initial eccentricity and orbital frequency, and $\sigma(e)$ is given in Eq.~(15) in~\cite{Moore:2018kvz}. The function $I(e)$ is given in closed form in terms of hypergeometric functions in Eq.~(58) in~\cite{Moore:2018kvz}, or in a post-circular expansion in Eq.~(59) therein. Each harmonic of these waveforms has qualitatively similar behavior to quasi-circular waveforms, solidifying the fact that eccentric orbits can be thought of as epicylces of circular orbits. This can be seen through the spectrograms in Fig.~(4) in~\cite{Moore:2018kvz}.

The post-Newtonian extension of the NeF waveform was considered in~\cite{Moore:2019xkm}. Much of the machinery developed for the NeF waveforms also applies at higher PN order. However, there is one critical feature that is present at higher post-Newtonian order, namely the splitting of harmonics. Periastron precession enters the post-Newtonian two-body problem at relative 1PN order, which causes eccentric orbits to be characterized by two frequencies: an azimuthal (orbital) frequency, and a radial frequency. Physically, this causes a Zeeman splitting-like effect of each orbital harmonic within the waveform, as can be seen by Fig.~(4) in~\cite{Moore:2019xkm}. This model has been shown to achieve matches greater than $0.97$ against numerical post-Newtonian waveforms up to eccentricities of $0.8$ for widely separated binaries. The maximum eccentricity at which it is valid decreases with decreasing separation due to issues related to the convergence of the PN sequence for eccentric systems.

\subsubsection{Effective fly-by approach}

A recent development in pursuit of analytic waveforms for highly eccentric binaries is the {effective fly-by} (EFB) formalism~\cite{Loutrel:2019kky}. The formalism seeks an accurate description of the gravitational wave bursts from individual pericenter passages, which may be strung together to create inspiral waveforms using the timing models described in Sec.~\ref{timing}. At leading post-Newtonian order, the relationship between the orbital phase and coordinate time is given by the transcendental Kepler's equation, which admits no known closed-form analytic solution. One circumvents this issue by using Kepler's equation to write functions that depend on the orbital phase as functions of time using Fourier series, as is done in Eq.~\eqref{eq:cos-fourier}-\eqref{eq:sin-fourier}. At low eccentricities ($e\ll1$), these series can be truncated at a small, finite number of terms to obtain an accurate representation of the orbit. However, at high eccentricities ($e\sim1$) this cannot be achieved and one would have to keep hundreds to thousands of terms to obtain the same accuracy.

At it's core, the formalism relies on a procedure to re-sum series of the structure in Eq.~\eqref{eq:cos-fourier}. Series that depend on Bessel functions of the form $J_{k}(k e)$ are known as Kapteyn series. In a select few cases, these series can be re-summed exactly, as was done in the quadrupole radiation calculations of Peters \& Mathews~\cite{Peters:1963ux}. In general, this is not the case and one has to rely on approximate methods to re-sum these series. The general procedure that is utilized in the EFB formalism was originally developed in~\cite{Loutrel:2016cdw} for use in the calculation of non-linear effects in the gravitational wave fluxes. The procedure follows a few basic steps:
\begin{enumerate}
\item Replace the Bessel function $J_{k}(k e)$ with it's uniform asymptotic expansion about $k\rightarrow\infty$. To leading order $J_{k}(ke) \sim K_{1/3}[(2/3)\zeta^{3/2}k]+ {\cal{O}}(1/k)$, where $K_{n}(x)$ is the modified Bessel function of the second kind, and $\zeta$ is a known function of the orbital eccentricity $e$ (see Eq.~(26) in~\cite{Loutrel:2019kky}). For a comparison of the asymptotic expansion to the exact function, see Fig.~2 in~\cite{Loutrel:2016cdw}.\\
\item Replace the sum over $k$ with an integral. There is a practical issue of whether or not the integral can be evaluated analytically. In the case of Eq.~\eqref{eq:cos-fourier}, this can be done once the integral is extended down to $k=0$.\\
\item Expand about high eccentricity, specifically $1 - e^{2} \ll 1$. The previous two steps are most accurate in the high eccentricity limit. The expansion also serves to speed up numerical computations, since the result of the previous step generally produces hypergeometric functions that are slow to evaluate.\\
\end{enumerate}
In the case of Eq.~\eqref{eq:cos-fourier}, this procedure produces
\begin{equation}
\cos\phi \sim -1 + \frac{2 \cosh[(1/3) \sinh^{-1}\psi]}{\sqrt{1 + \psi^{2}}} + {\cal{O}}(1-e^{2})\,,
\end{equation}
where $\psi = (3/2)\ell/\zeta^{3/2}$. The time-domain quadrupole waveform can then be written purely in terms of $\psi$ and is given in Eq.~(45) in~\cite{Loutrel:2019kky}.

The above re-summation procedure solves the problem of the conservative dynamics of the binary at leading post-Newtonian order. Using the balance laws in the quadrupole approximation, one can obtain evolution equations for the semi-latus rectum $p$ and orbital eccentricity $e$ in the adiabatic limit (see Eqs.~(36)-(37) in~\cite{Loutrel:2019kky}). To obtain an approximate solution to these coupled equations near pericenter, the EFB formalism relies on a simple Taylor series expansion of the form
\begin{equation}
\label{eq:taylor}
\mu^{a} = \mu^{a}_{0} + \left(\frac{d\mu^{a}}{d\ell}\right)_{0} \ell(t) + {\cal{O}}[\ell(t)^{2}]
\end{equation}
where $\mu^{a}$ is either $p$ or $e$, and the underscore designates the value of the relevant quantity at $\ell = 0$. Once radiation reaction is considered, the mapping between $\ell$ and coordinate time $t$ is no longer linear, and one has to solve the differential equation $d\ell/dt = n[p(\ell), e(\ell)]$. Making use of the Taylor series in Eq.~\eqref{eq:taylor}, this equation can be solved to obtain $\ell(t)$ given in Eq.~(44) in~\cite{Loutrel:2019kky}. A comparison of these analytic function to the numerical integration of the quadrupole radiation reaction equations can be found in Fig.~2 of~\cite{Loutrel:2019kky}. The time domain radiation reaction model, and the gravitational wave polarizations combined with the re-summed conservative dynamics constitute the EFB-T model.

Validation of the EFB-T model were also considered in~\cite{Loutrel:2019kky}. The match statistic was used to quantify the faithfulness of the model against numerical leading PN order quadrupole waveforms, which was found to be above 0.97 for much of the parameter space of relevance to ground-based detectors. While this is useful in quantifying the errors associated with the approximations used to develop the model, an important question is whether or not the model accurately captures a burst from an eccentric binary that we might expect in nature. To do this, the EFB-T model was compared to several numerical relativity fly-by waveforms computed in~\cite{East:2011xa}. For pericenter distances of $r_{p} > 10 M$, the match is greater than 0.94, but peaks at different orbital parameters than those used in the numerical relativity simulations, indicating systematic bias as a result of the limited leading post-Newtonian order approximation used to develop the model. For smaller pericenter distance, the match decreases rapidly, owing to the fact that nonlinear effects break the symmetrical structure of the burst in the numerical relativity simulations (see Fig.~6-8 in~\cite{Loutrel:2019kky}). 

There is still much open work that remains to be done in the EFB formalism. So far, studies have focused on the leading post-Newtonian order description of both the conservative and dissipative dynamics of the binary. Beyond this order, the conservative dynamics becomes significantly more complicated, and its not clear if the same re-summation techniques can be applied. A more tantalizing question is whether this approach can be applied to effective description of the two-body problem, such as the effective Kerr spacetime in Sec.~\ref{kerr} or the effective one-body formalism. Further, the modeling currently only suffices for test particles, which may be a good approximation for black holes, but will fail to capture the rich physics that is present if at least one of the binary components is a neutron star.

\subsubsection{Timing models}
\label{timing}

One of the possible detection strategies for gravitational wave bursts from highly eccentric binaries is power stacking, discussed in Sec.~\ref{stack}. However, the issue with this method is distinguishing the bursts from glitches, anomalous noise sources in ground-based detectors, many of which resemble the time and frequency morphology of the bursts. Thus, this detection method requires a means to distinguish the bursts from glitches. The key is to realize that eccentric binaries will emit a burst with each pericenter passage as it inspirals, and that the sequence of bursts will follow a predictable track in time-frequency space, whereas glitches are generally random events. The goal is then to use our knowledge of the two-body problem in general relativity to construct a timing model (called a burst model in earlier literature), for the time-frequency track of the inspiral sequence~\cite{Loutrel:2014vja, Loutrel:2017fgu}.

\begin{figure}
\begin{centering}
\includegraphics[scale=0.3]{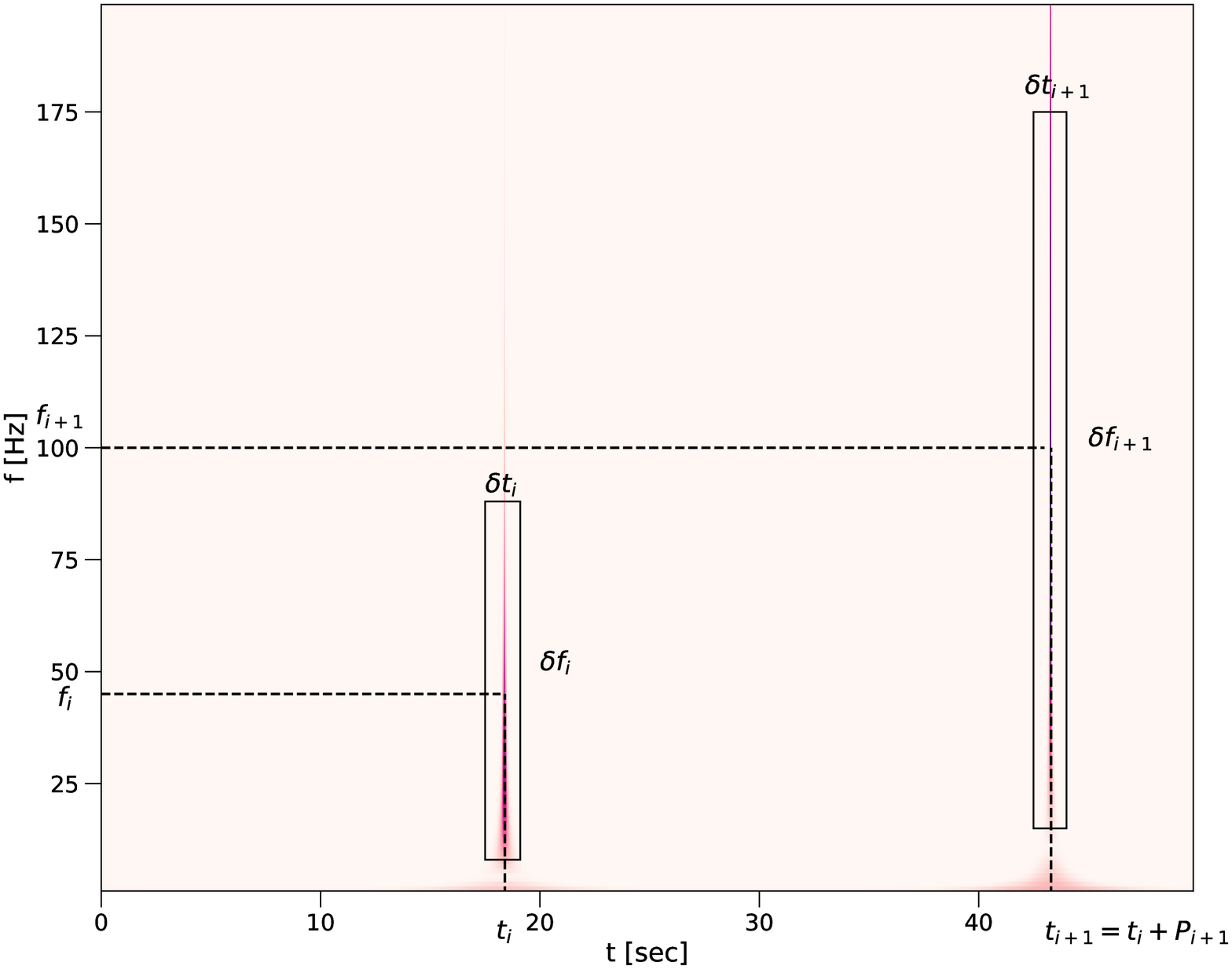}
\caption{\label{schem} Schematic of the timing model for bursts emitted during the inspiral of a highly eccentric binary. In the simplest model, each burst is described by a box characterized by a centroid with frequency $f_{i}$ and time $t_{i}$, and widths $(\delta t_{i}, \delta f_{i})$. The time between bursts is simply the orbital period of the binary, while the central frequency and widths are related to the properties of the power spectrum of each burst.}
\end{centering}
\end{figure}

The starting point to construct a {timing model} is a visualization of the burst in time-frequency space, see for example Fig.~\ref{schem}. The burst are effectively two dimensional shapes in this space, so a first approximation to this is to characterize the bursts by rectangles in time and frequency. One could choose a more apt shape, such as an ellipse, but this can be achieved by a simple re-interpretation of the widths of rectangles. Each rectangle will then be characterized by a centroid $(t_{i}, f_{i})$ and widths $(\delta t_{i}, \delta f_{i})$ (or in the case of ellipses, the major and minor axes). To construct a timing model of the bursts, three ingredients are required:
\begin{enumerate}
\item Oribtal evolution: A model that determines the orbital parameters of one orbit from those of the previous orbit under the effect of radiation reaction.\\
\item Centroid model: A model that determines the centroid of one burst $(t_{i}, f_{i})$ from the centroid of the previous burst $(t_{i-1}, f_{i-1})$.\\
\item Volume model: A model that determines the extent $(\delta t_{i}, \delta f_{i})$ of each burst in time and frequency.\\
\end{enumerate}

First, consider the orbital evolution. In the time domain, the orbital evolution is described by a step like behavior near pericenter, due to the rapid emission of gravitational waves during pericenter passage. This can be approximated by treating the orbital parameters as constant throughout the orbit, but allowing them to change instantaneously at pericenter. The orbit in between consecutive pericenter passages is described by two parameters, namely the orbital energy $E$ and orbital angular momentum $L$. The change in orbital energy and angular momentum due to radiation reaction is dependent on these parameters, as well as the orbital phase, namely $\dot{E} = \dot{E}(E, L, \phi)$ and $\dot{L} = \dot{L}(E, L, \phi)$. From these, one must compute the change in orbital energy and angular momentum over the orbit. This can be achieved via the adiabatic approximation, where the average of these rates of change allows one to directly compute the change in these quantities if one knows the orbital period. Thus, the mapping between the orbital periods of one orbit and the previous orbit is
\begin{align}
\label{eq:E-next}
E_{i} &= E_{i-1} + \langle \dot{E} \rangle (E_{i-1}, L_{i-1}) P_{i-1}\,,
\\
\label{eq:L-next}
L_{i} &= L_{i-1} + \langle \dot{L} \rangle (E_{i-1}, L_{i-1}) P_{i-1}\,,
\end{align}
where $\langle \dot{E} \rangle$ and $\langle \dot{L} \rangle$ are the orbit averaged energy and angular momentum fluxes due to gravitational waves, and $P$ is the orbital period. These mappings are true regardless of the approximations that one applies to the orbital dynamics. However, these quantities have so far only been computed within the post-Newtonian formalism, using the {quasi-Keplerian} description of the orbit. With this description, one can alternatively work in terms of the pericenter velocity $v_{p}$ and time eccentricity $e_{t}$, which are known functions of the orbital energy and angular momentum. The mappings $v_{p,i}(v_{p,i-1}, e_{t,i-1})$ and $e_{t,i}(v_{p,i-1}, e_{t,i-1})$ are given in a post-Newtonian expansion about $v_{p,i-1} \ll 1$ in Eqs.~(53)-(59), (71)-(74), (88)-(95), \& (107)-(112) in~\cite{Loutrel:2017fgu}.

Now consider the centroid mapping. Similar arguments regarding the step like behavior of the orbital evolution can be applied to write the time mapping purely in terms of the orbital period, specifically
\begin{equation}
t_{i} = t_{i-1} + P_{i}\,.
\end{equation}
Just like the mappings in Eq.~\eqref{eq:E-next}-\eqref{eq:L-next}, this applies to any generic orbital description. Within the quasi-Keplerian formalism, the orbital period is a known function of the orbital energy and angular momentum, or alternatively, the pericenter velocity and time eccentricity. The mapping $P_{i}(v_{p,i-1}, e_{t,i-1})$ is given in Eq.~(62), (77), (97), \& (118) in~\cite{Loutrel:2017fgu}. The frequency mapping, on the other hand, requires knowledge of the power spectrum of the gravitational wave burst. However, from simple scaling arguments, one might expect the peak frequency of the power spectrum to scale inversely with the pericenter timescale $\tau$, namely
\begin{equation}
f_{i} = \frac{1}{2\pi \tau_{i}}\,,
\end{equation}
where
\begin{equation}
\tau = \frac{\text{pericenter distance}}{\text{pericenter velocity}}\,.
\end{equation}
At leading post-Newtonian order, calculations of the power spectrum of bursts from parabolic orbits have provided evidence of this (see Fig.~(7) of~\cite{Turner}), but this argument likely holds for generic orbital descriptions. The mapping $f_{i}(v_{p,i-1}, e_{t,i-1})$ is given in Eq.~(61), (80), (98), \& (124) in~\cite{Loutrel:2017fgu} for the quasi-Keplerian formalism.

Finally, consider the volume mapping, which is characterized by the widths $\delta t_{i}$ and $\delta f_{i}$. The models developed in~\cite{Loutrel:2014vja,Loutrel:2017fgu} choose the widths to be proportional to the pericenter timescale and peak frequency of the burst, specifically
\begin{align}
\delta t_{i} &= \xi_{i} \tau_{i}\,,
\\
\label{eq:f-width}
\delta f_{i} &= \xi_{f} f_{i}\,.
\end{align}
The proportionality factors $(\xi_{t}, \xi_{f})$ must be chosen from data analysis considerations. One possible way of fixing them is to choose their values such that a certain percentage of the power in each burst is contained in the box characterized by $(\delta t_{i}, \delta f_{i})$. Alternatively, if one wanted to associate the bursts with sine-gaussian wavelets, the proportionality factors are related to the Q-factor of the wavelet.

The entire sequence of N bursts is characterized by the set $\{t_{i}, f_{i}, \delta t_{i}, \delta f_{i}| i=1,...,N\}$. Naively, one might expect that $4N$ parameters are need to describe the entire track of the bursts in time-frequency space. However, the conservative and dissipative dynamics of the binary are only described by four parameters (neglecting spins), specifically $(m_{1}, m_{2}, E_{0}, L_{0})$ where $m_{1,2}$ are the component masses, and $(E_{0}, L_{0})$ are the orbital energy and angular momentum in the first burst in band of the gravitational wave detector. From Eqs.~\eqref{eq:E-next}-\eqref{eq:f-width}, the entire track of the bursts in time-frequency space is determined by only these four parameters. As a result, these four parameters can be determined by a finite number of bursts in principle, and neglecting statistical uncertainties. Since the widths are determined by $\tau$, each burst only actually provides at most two independent parameters. Thus, at the very least, one needs only two bursts to invert the parameters $(t_{i}, f_{i}, \delta t_{i}, \delta f_{i})$ to $(m_{1}, m_{2}, E_{0}, L_{0})$.

While the timing model reviewed here applies in any generic orbital description, it has only been computed to third post-Newtonian order in the quasi-Keplerian formalism. Further, the model has only been validated against numerical integration of the post-Newtonian equations of motion. It has not yet been validated in a mock data analysis study. It is also worth noting that these timing models don't just apply for power stacking searches, but can also be used in matched filtering searches using the EFB waveforms. In~\cite{Loutrel:2019kky}, it was shown that the EFB waveforms combined with a leading post-Newtonian order timing model achieves a high match against a numerical leading post-Newtonian order quadrupole waveform. However, if a small error is introduced into the timing model, the match drops off rapidly, and one could potentially lose detection of the bursts. One would thus likely need to work beyond the post-Newtonian formalism to obtain the most accurate timing model compared to binaries that we would observe in nature.

\subsection{Tests of general relativity}

One of the most interesting applications of eccentric binaries is as a means of {testing general relativity}. As the eccentricity increases, the velocity at closest approach also increases up to approximately $41\%$ from the quasi-circular case. Further, the innermost stable orbit, which one may consider the end of the inspiral phase of the coalescence, is generically smaller for eccentric systems. In the Schwarzschild spacetime, the innermost stable circular orbit corresponds to $r=6M$, where $M$ is the mass of the black hole. For eccentric orbits, the smallest the pericenter distance can be is $4M$ for bound orbits, corresponding to a pericenter velocity of $v_{p} \sim 0.7 c$, where $c$ is the speed of light in vacuum.

These considerations are important when considering alternative theories of gravity. Orbital dynamics in many of these theories can be considered as deformations of the dynamics within general relativity, specifically the corrections to orbital quantities scales with the characteristic velocity of the system. As an example, consider the correction to the orbital energy in {Einstein-dilaton-Gauss-Bonnet gravity}, a theory of gravity that modifies general relativity through the introduction of quadratic curvature invariants in the action. The orbital energy and angular momentum in this theory are~\cite{Loutrel:erratum}
\begin{align}
E = E_{\rm GR} \left[1 + \frac{\zeta}{6} \left(\frac{1-e}{1+e}\right) \left(\frac{M}{r_{p}}\right)^{2}\right]
\\
L = L_{\rm GR} \left[1 - \frac{\zeta}{12} \frac{3+e^{2}}{(1+e)^{2}} \left(\frac{M}{r_{p}}\right)^{2}\right]
\end{align}
where $\zeta$ is the coupling constant of the theory, and $M/r_{p} \sim v_{p}^{2}$ by the viral theorem. These expressions also illustrate an important point, namely that corrections to orbital dynamics in theories of modified gravity are eccentricity dependent, and as a result, constraints on the coupling constants of the theory from parameter estimation will also depend on the eccentricity of the binary.

Projected constraints on example modified theories have been considered in both the low and high eccentricity limits. In the low eccentricity limit, waveforms have been computed in {Jordan-Brans-Dicke-Fierz theory} using the stationary phase and post-circular approximations~\cite{Ma:2019rei}. This theory reduces to general relativity in the limit $\omega \rightarrow \infty$, with $\omega$ the coupling constant. With analytic waveforms, the projected constraints on $\omega$ were obtained using a Fisher analysis, an analytic parameter estimation technique valid in the high SNR limit and when the detector noise is stationary and gaussian. Ma \& Yunes~\cite{Ma:2019rei} found that as the orbital eccentricity increases from zero (the quasi-circular limit), the constraint on $\omega$ initially deteriorates, but recovers for eccentricities above $0.03$ (see Fig.~1 therein). Similar non-trivial behavior of the constraint with increasing eccentricity was found by Moore \& Yunes in~\cite{Moore:2020rva}, who studied constraints on {Brans-Dicke theory} and Enstein-dilaton-Gauss-Bonnet gravity. Using Markov-Chain Monte Carlo methods, Moore \& Yunes found that the constraints are strongest for moderately eccentric systems with $e\sim0.4$, with the projected constraints being an order of magnitude better than current constraints with quasi-circular binaries.

In the high eccentricity limit, constraints on modified theories were considered with the timing models considered in the previous section in~\cite{Loutrel:2014vja}. At the time, analytic waveforms for the gravitational wave bursts from highly eccentric systems were not available, but one could compute the timing model for generic deformations from general relativity to develop a parameterized post-Einsteinian (ppE) model. To obtain projected constraints, the authors considered variations of time and frequency centroid mappings with respect to the ppE parameters and required them to be at least as large as variations of the sequence with respect to the parameters within general relativity (the masses, orbital eccentricity, etc.), i.e.
\begin{equation}
\label{eq:time-var}
\frac{\partial (\Delta t, \Delta f)}{\partial \lambda_{\rm ppE}^{a}} \delta \lambda_{ppE}^{a} > \frac{\partial (\Delta t_{\rm GR}, \Delta f_{\rm GR})}{\partial \lambda_{\rm GR}^{a}} \delta \lambda_{\rm GR}^{a}\,.
\end{equation}
Constraints on two example theories were considered in~\cite{Loutrel:2014vja}, namely Einstein-dilaton-Gauss-Bonnet and Brans-Dicke gravity. In both of these theories, the leading order corrections to the timing model come from the existence of scalar {dipole radiation}, which are $-1$PN order effects on the flux of energy and angular momentum from gravitational waves. Making use of Eq.~\eqref{eq:time-var}, the projected constraints on the coupling constants for these theories were found to be comparable, but not better than, astrophysical constraints at the time. These upper limits are strictly estimates, since there are currently no waveforms for highly eccentric binaries in modified theories for use in parameter estimation studies.

\section{Binaries with Neutron Stars}
\label{ns}

So far, the discussion in this chapter has focused on black holes, which can be modeled as effective point particles in post-Newtonian theory and beyond. However, several studies have considered the possibility of at least one of the compact objects comprising an eccentric binary being a neutron star. In order to effectively describe such a system, one must consider the effects of matter on these systems. This chapter will review some of the most recent findings regarding neutron stars in eccentric binaries.

\subsection{Tidal interactions and resonances}
\label{tide}

In~\cite{Stephens:2011as}, {neutron stars} with polytropic equations of state were considered in the context of highly eccentric {black hole-neutron star binaries} using numerical relativity simulations. Effectively, the compact objects were setup on unbound, nearly parabolic orbits with varying pericenter distance $r_{p}$ (or alternatively, impact paramters). Due to the limitations of computational techniques at the time, the simulations were limited to $r_{p} < 15 M$. Three possible scenarios emerged depending on the value of the initial pericenter separation: a direct plunge, an initial burst leading into a highly eccentric orbit followed by a plunge at the next pericenter passage, or a single pericenter passage leading to a highly eccentric orbit that does not merge. The latter of these could not be tracked beyond the initial fly-by due to computational limitations. A discussion of the end state of the plunge will be presented in Sec.~\ref{post-merger}. In the latter two cases, extraction of the gravitational waves in the simulation found fundamental oscillations in the waveform after the burst associated with the initial fly-by, which are not present in the binary black hole scenario (see the top panel of Fig.~(3) in~\cite{Stephens:2011as}). The oscillations are a result of the fundamental modes (f-modes) of the neutron star, and indicate the presence of rich dynamics in such binaries. 

To get a better understanding, it is useful to consider analytic calculations of perturbations of neutron stars. In the context of binary systems, to leading order the neutron star is acted upon by a tidal potential, generated by the second compact object, of the form
\begin{equation}
U_{\rm tidal} = \frac{1}{2} {\cal{E}}_{ij} x^{i} x^{j}
\end{equation}
where ${\cal{E}}_{ij}$ is the electric part of the tidal quadrupole moment~\cite{Poisson:2009qj}. The tidal potential acts to create a displacement of the fluid elements $x^{i} \rightarrow x^{i} + \xi^{i}$, which deforms the quadrupole moment of the star. In a spherical harmonic decomposition, the time dependent amplitude of the quadrupole moment obeys
\begin{equation}
\ddot{Q}_{m} + \gamma \dot{Q}_{m} + \omega^{2} Q_{m} = W_{2m} K_{2m} \frac{m_{2}}{r^{3}} e^{-im\phi}
\end{equation}
where $W_{lm}$ is given in Eq.~(24) in~\cite{Press:1977}, $K_{lm}$ is an equation of state dependent parameter given in Eq.~(38) of~\cite{Press:1977}, $m_{2}$ is the mass of the other body in the binary, $r$ is the orbital separation, and $\phi$ is the orbital phase. The above equation describes the $l=2$ {f-mode} of the neutron star, with damping coefficient $\gamma$ and frequency $\omega$. 

In the quasi-circular limit, the orbital separation $r$ is a slowly evolving quantity under radiation reaction, and thus the tidal perturbation of the neutron star effectively becomes fixed, with its adiabatic evolution becoming subdominant. This is the so-called {static tide}. However, this approximation does not hold for the entire coalesence, the last few orbit of which are known to be highly dynamical, leading the tide to become highly time dependent and to the excitation of f-modes. For eccentric systems, especially those with $e\sim1$, the situation is entirely reversed. The rapid nature and small orbital separation at pericenter passage lead to the tidal potential being it strongest during closest appoach, causing the excitation of f-modes at each passage. Energy and angular momentum are deposited in the modes from the orbital energy and angular momentum, which in turn modifies the inspiral rate beyond that of the typical radiation reaction effects of point particles. Further, the f-mode leads to a time dependent quadrupole moment for the neutron star, generating additional gravitational waves from the system through Eq.~\eqref{eq:hij}.

Since the studies of~\cite{Stephens:2011as, East:2011xa, Gold:2011df}, there has been much work on trying to understand these tidal effects in more detail, primarily through the use of analytic and numerical calculation in the post-Newtonian approximation. Yang et. al.~\cite{Yang:2018bzx} performed detailed calculations of the energy and angular momentum deposited into the f-modes in eccentric binaries, and compared the analytic post-Newtonian predictions to those of numerical relativity simulations. Both methods found that typically $10-20\%$ of the energy and angular momentum in gravitational waves from a single fly-by can be deposited into the modes, although there is a disparity between the two techniques, which is a result of the limited accuracy of the post-Newtonian approximation as well as numerical error in the simulations. 

These results were extended in~\cite{Yang:2019kmf} to create a model of the evolution of the orbit under tidal effects. One of the main challenges in developing such a model for arbitrary eccentricity is the lack of closed-form expressions for $(r,\phi)$ as functions of time (see Sec.~\ref{dyn}). However, this was solved by the introduction of Hansen coefficients, which were only able to be evaluated in the low eccentricity limit. Despite this, Yang found that the phase difference in the orbital evolution between including and neglecting tidal effects to only be $\sim0.06$ radians for a $(1.3,1.3) M_{\odot}$ binary neutron star with $e=0.2$ at orbital frequency of 50 Hz, which is only detectable for loud events with third generation detectors. 

Yang~\cite{Yang:2019kmf} also explored the possibility of {tidal resonances} in {binary neutron stars}, where the f-mode frequency is an integer multiple of the orbital frequency. In such a scenario, the resonance causes a larger amount of orbital energy and angular momentum to be deposited into the mode, enchancing the phase shift in the orbital evolution. Generally, the f-mode frequency is significantly larger than the orbital frequency in such a binary, while the effects of resonances with large harmonic number are subdominant. However, in the late stages of the inspiral, it is possible for resonances to be achieved with the $k=2$ and $k=3$ harmonics, which Yang argued were the dominant contribution to the phase shift of the binary and the only ones of relevance to low eccentricity binaries. For the same $(1.3,1.3) M_{\odot}$ binary neutron star, the phase shift due to a $k=3$ resonance was found to be $\sim0.5$ radians. Despite, the roughly order of magnitude increase in the phase shift from the non-resonant case, such an effect is still of most relevance to third generation detectors. These estimates, in both the resonant and non-resonant cases, do not take into account possible de-phasing at lower frequency, which would correspond to larger eccentricities.

Vick and Lai~\cite{Vick:2018, Vick:2019cun} have investigated these effects in the context of highly eccentric binaries using numerical integrations of the post-Newtonian equations of motion. Generically, in the non-resonant case they found that the de-phasing in the gravitational waves has a non-trivial dependence on the eccentricity, which can be seen in Fig.~7 in~\cite{Vick:2019cun}. Further, the cumulative phase shift in the gravitational waves was found to be larger than the results of Yang for smaller initial pericenter distances, regardless of the value of the eccentricity. In the resonant case, the phase shift due to the resonance was found to always be less than 0.1 radians, indicating that tidal resonances are not the dominant contribution to the de-phasing.

Vick and Lai~\cite{Vick:2018} also found a third plausible scenario of relevance to mode excitation in highly eccentric binaries. Recall that for parabolic orbits that the binding energy is zero, i.e. $\epsilon = 0$, while for bound eccentric orbits $\epsilon < 0$. Thus, for highly eccentric binaries with $e\sim1$, $\epsilon\sim0$. In such a scenario, it was found that the energy deposited into the mode is comparable to or larger than the binding energy, causing the mode to grow chaotically (see the bottom panel of Fig.~3 in~\cite{Vick:2018} and Fig.~3 in~\cite{Vick:2019cun}). From Fig.~2 in~\cite{Vick:2018}, the amount of energy extracted from {chaotic modes} over multiple orbits is typically several orders of magnitude larger than both the non-resonant and resonant cases. Under the effects of radiation reaction, Vick and Lai found that the chaotic modes will not grow continuously, but will eventually settle into a quasi-steady state as the binary loses eccentricity.

The authors of these studies argue that these effects can be used to constrain the equation of state of neutron stars. To date, there are no studies that provide predictions for constraints on equations of state for highly eccentric neutron star binaries using parameter estimation with current and future ground based detectors.

\subsection{Post-merger and remnants}
\label{post-merger}

{Numerical relativity} simulations of the merger phase and remnants of the coalescence for highly eccentric binaries with neutron stars have been considered in~\cite{Stephens:2011as, East:2011xa, Gold:2011df, East:2012ww, East:2015yea, East:2015vix, East:2016zvv}. The earliest study focused on black hole neutron-star binaries with a mass ratio of 4:1. The ultimate fate of the neutron star was found to be dependent on the initial pericenter separation of the dynamical capture and on the equation of state, which can be seen from Fig.~1 in~\cite{Stephens:2011as}. The first possible scenario, for $r_{p} \sim 5M$, is a direct plunge of a highly deformed neutron star with little tidal disruption, resulting in less that $1\%$ of the rest mass available to create an {accretion disk} post-merger. For slightly larger periapsis, $r_{p} \sim 6.81M$, the neutron star is disrupted into a long tidal tail, leading to an accretion disk with $\sim12\%$ of the rest mass at late times. At larger pericenter distances, mass transfer between the neutron star and black hole can occur during initial passage, with a subsequent highly distorted neutron star, ultimately leading to the creation of a nascent disk at late times.

A follow up to this study was performed in~\cite{East:2011xa}, including the effects of black hole spin. Generically, the {innermost stable orbit} (ISO) is dependent on whether the black hole spin is aligned or anti-aligned with the orbital angular momentum, leading to prograde and retrograde orbits respectively. In the prograde (aligned) case, the ISO is closer to the black hole, and the neutron star experiences large tidal forces before merger, leading to the formation of accretion disks with larger rest mass compared to the non-spinning and retrograde cases. In the retrograde case, the ISO is located farther from the black hole, and tidal disruption is not as efficient, with accretion disks containing less than $1\%$ of the neutron star's rest mass. The equation of state also has a significant impact on the post-merger result, with more compact stars experiencing less tidal disruption and resulting in accretion disks with smaller rest mass.

In~\cite{East:2012ww}, East \& Pretorius considered the case of non-spinning binary neutron star mergers resulting from dynamical captures. The neutron stars were modeled using piecewise polytropic equations of state, and thus mangetohydrodynamic and neutrino cooling effects were neglected. As with the black hole-neutron star case, the final remnant depends sensitively on the initial pericenter distance. For exceptionally small periapsis ($r_{p} \le 5M$), the neutron stars directly collapse into a single black hole after collision, due to the lack of sufficient angular momentum to support a hypermassive remnant. The black hole mass is typically $>0.98 M$ with spin $\chi>0.5$. For larger periapsis, the collision of the neutron stars forms a deformed hypermassive neutron star supported by angular momentum. The hypermassive neutron star radiates gravitational waves due to its deformed quadrupole moment, losing angular momentum in the process and ultimately collapsing into a black hole. 

Full general relativistic hydrodynamic simulations of {hypermassive neutron stars} resulting from mergers in the dynamical capture scenario have been considered in~\cite{East:2015vix, Paschalidis:2015mla, East:2016zvv}. Paschalidis et. al.~\cite{Paschalidis:2015mla} found that the hypermassive neutron stars form a {one arm spiral instability}, an $m=1$ spherical harmonic mode, which was the first time the instability had been observed in binary neutron star mergers. In East et. al.~\cite{East:2015vix}, it was found that the instability was most pronounced for binary neutron stars with small, but non-negligible spin. Further, in~\cite{East:2016zvv}, it was found that the equation of state strongly influences the relative amplitudes of $m=2$ and $m=1$ azimuthal modes, with stiffer equations of state having stronger $m=2$ modes and weaker $m=1$ modes. The instability is important since it contributes to the post-merger gravitational waves emitted by the hypermassive neutron star. Generically, the post-merger signal is dominated by the $l=m=2$ modes of the neutron star, but these occur at high frequencies of a few kilohertz. The $l=2$, $m=1$ mode generated by the instability has roughly half the frequency of the dominant mode, and thus may be easier to detect by ground based detectors.

When considering the mergers of neutron stars, it is also important to characterize the amount of material that is ejected from the system, which can generate electromagnetic counterparts in the form of {kilonovae}. East \& Pretorius~\cite{East:2012ww} also considered the amount of ejecta created in the merger of their simulations, and found that the mass is at most a few percent of a solar mass at mildly relativistic velocities, specifically $10-30\%$ the speed of light. These results were later extended to the case of spinning neutron stars in~\cite{East:2015yea}, where it was found that the ejecta mass increases with increasing spin, up to a third of a solar mass for rapidly rotating neutron stars (see the bottom left panel of Fig.~4 therein).

\section{Cross-References}

Below is a list of chapters with related topics to the discussion in this chapter.\\

\noindent \textit{Introdcution to gravitational wave astronomy} by Nigel Bishop\\
\textit{Binary neutron stars} by Luca Baiotti\\
\textit{Black hole-neutron star mergers} by Francois Foucart\\
\textit{Dynamical formation of stellar-mass binary black holes} by Bence Kocsis\\
\textit{Formation channels of single and binary stellar-mass black holes} by Michela Mapelli\\
\textit{Post-Newtonian templates for gravitational waves from compact binary inspiral} by Soichiro Isoyama, Hiroyuki Nakano, and Riccardo Sturani\\
\textit{Numerical relativity for gravitational wave source modelling} by Zhoujian Cao\\
\textit{Gravitational waves in alternative theories of gravity} by Mariafelicia De Laurentis and Ivan De Martino\\
\textit{Strong-field tests of gravity with gravitational waves} by Kent Yagi\\
\textit{Principles of gravitational wave data analysis} by Andrzej Krolak\\

%
%
%
%

\end{document}